\documentclass{aastex63}

\newcommand\lya{$\mathrm{Ly}\,\alpha$}

\newcommand\targ{HIP~67522} 

\graphicspath{{./}{figures/}}

\received{\today}
\revised{\today}
\accepted{\today}
\submitjournal{ApJ}

\shorttitle{Young Sun HIP 67522}
\shortauthors{Froning et al.}

\begin{document}


\title{X-ray and UV Observations of the Young Sun HIP~67522: Evidence of Lyman-alpha Absorption Within the Planetary System}

\correspondingauthor{Cynthia S. Froning}

\author[0000-0001-8499-2892]{Cynthia S. Froning}
\affiliation{Southwest Research Institute, San Antonio, TX 78238}

\author[0000-0002-1176-3391]{Allison Youngblood}
\affil{Goddard Space Flight Center, Greenbelt, MD 20771}

\author[0000-0001-9667-9449]{David J. Wilson}
\affil{Laboratory for Atmospheric and Space Physics, University of Colorado, 600 UCB, Boulder, CO 80309}

\author[0000-0002-7119-2543]{Girish M. Duvvuri}
\affil{Vanderbilt University, Nashville, TN 37240}

\author[0000-0002-1002-3674]{Kevin France}
\affil{Department of Astrophysical and Planetary Sciences, University of Colorado, Boulder, CO 80309, USA}
\affil{Laboratory for Atmospheric and Space Physics, University of Colorado, 600 UCB, Boulder, CO 80309}

\author[0000-0002-5094-2245]{P.\ Christian Schneider}
\affil{Hamburger Sternwarte, Gojenbergsweg 112, 21029 Hamburg }

\author{J. Sebastian Pineda}
\affil{Department of Astrophysical and Planetary Sciences, University of Colorado, Boulder, CO 80309, USA}

\author{Alexander Brown}
\affil{Department of Astrophysical and Planetary Sciences, University of Colorado, Boulder, CO 80309, USA}

\author[0000-0003-1133-1027]{Angeli Sandoval}
\affil{Department of Astrophysics, The Graduate Center of the City University of New York, 365 Fifth Avenue, New York, NY 10016, USA}

\author[0000-0002-1242-5124]{Thomas Ayres}
\affil{Department of Astrophysical and Planetary Sciences, University of Colorado, Boulder, CO 80309, USA}

\begin{abstract}
We present ultraviolet (UV) spectroscopy of the 17 Myr, G0V star, \targ. The UV spectrum is characterized by strong chromospheric and transition region emission lines. There was moderate spectral variability during the observations consisting of 15\%  stochastic fluctuation and two small flares releasing  $E_{UV} \simeq 2-4\times10^{32}$~ergs in each event. 
We compare the broadband spectral energy distribution (SED; 4.7~\AA\ -- 13.0~$\mu$m) of the star first presented in \citet{thao24-1} to the solar SED and show that X-ray/UV (XUV) flux density at 1~AU is $10^{2}-10^{5}$ stronger (from 1000 ~\AA\ down to 5~\AA) in the young star compared to the present-day Sun. Attempts to reconstruct the intrinsic \lya\ emission of the star failed to return physically realistic results. The discrepancy appears to arise from a population of neutral hydrogen within the system itself, extending to $> \pm500$~km~s$^{-1}$. The absorption could be due to outflow from exoplanet atmospheric loss or from a stellar component; such a picture would require high spectral resolution observations and/or UV transit spectroscopy to confirm. Finally, we examine the evolution of the XUV emission from solar-type stars from ages of 17~Myr to 9.4~Gyr and derive a scaling relation between FUV \lya\ and EUV emission as a function of stellar age. X-ray (1--100~\AA) and EUV (100--900~\AA) contributions to high energy emission are 329 and 672~ergs~cm$^{-2}$~s$^{-1}$ at 1~AU, respectively, suggesting that both may contribute to exoplanet heating at this epoch. The XUV emission levels at 17~Myr combined with the low density of the planet HIP67522b are consistent with models that predict that solar type stars born with high rotation and activity levels will drive substantial heating and escape on close-in, gaseous planets.

\end{abstract}


\section{Introduction} \label{sec:intro}

Analysis of exoplanet atmospheres depends on a strong understanding of the properties of the host star. The X-ray/ultraviolet (XUV) emission in particular governs planetary atmospheric heating and photochemistry. Both soft X-ray and EUV photons heat the planet atmosphere and can drive hydrodynamic escape \citep{lammer03-1,jackson12-1}. Extreme ultraviolet (100-911~\AA) photons are deposited in the upper atmosphere of the planet and are the dominant source of heating and escape for most planets. However, for close-in planets ($<$0.1~AU), the X-ray contribution to escape may be the primary driver at early epochs \citep{owen12-1,king21-1}.
X-ray emission peaks in young stars ($<$100~Myr), with $\sim$30~\% of the lifetime emission being radiated during the saturated phase, driven by a hot coronal component ($T_{hot} -23-30$~MK) component that declines very quickly \citep[$T_{hot} \propto age^{-0.3}$;][]{jackson12-1,gudel97-2}. Zero age main sequence (ZAMS) solar type stars can have a wide range of rotation rates and activity levels, with X-ray luminosities $\simeq1000\times$ that of the present-day Sun, converging to common levels within 2~Gyr \citep{ayres97-1,tu15-1,johnstone21-1}. The EUV emission at early epochs is less well understood due to the paucity of data in that waveband; the only previous astronomy mission to observe in this waveband was EUVE \citep{bowyer91-1}; the youngest star it observed was the $\sim$100~Myr star, EK~Dra  \citep{gudel97-1,ribas05-1}. 

XUV radiation has a particularly strong effect on low mass planets (super-Earths to sub-Neptunes) with low densities/large radii. The combination of the most intense phase of XUV emission occurring at early times and the fact that recently formed planets have radii inflated by residual heat from the star formation phase results in the dominant phase of hydrodynamic escape occurring in the first few hundred Myr \citep{tian08-1,owen13-1}. Models of XUV and planetary thermal evolution predict that there is a photoevaporation threshold wherein planets beyond a critical input radiative flux and density will not retain their hydrogen/helium envelopes. For the low mass planets, this may be the dominant cause of the gap in the planetary mass-orbital period plane:  planets further from the star retain their gas giant atmospheres, while close-in planets become rocky with partially stripped cores, as the latter evolve from sub-Neptune to super-Earth planets in the face of XUV energy deposition in early times \citep{fulton17-1,owen17-1,kubyshkina19-1}.

Given the importance of the early XUV, there have been numerous studies of the evolution of stellar high energy emission. 
For solar analogues, the benchmark reference has been the ``Sun in Time" paper presented by \citet{ribas05-1}. Based on observations of six solar-type stars plus the Sun, Ribas et al.\ (hereafter R05) calculated X-ray, EUV\footnote{Four of their stars had direct EUVE data, with the remainder interpolated from X-ray data.}, and FUV integrated fluxes and emission lines from 100~Myr to 6.7~Gyr. Since that time, the explosion in exoplanet atmospheric studies using transit spectroscopy plus the recognized importance of accurate high energy inputs of planets to interpret these data has vastly expanded the archive of high-quality X-ray and UV spectra of main sequence dwarf stars \citep{franceetal18-1,johnstone21-1,brown23-1,zhu25-1}. 

Our group has undertaken a large program using HST, XMM, and Chandra to acquire XUV spectra and time variability monitoring of 22 exoplanet host stars whose planet transits were observed in JWST Cycle 1. One target of interest is the young solar analogue, \targ. The star is a member of the 10--20~Myr old Sco-Cen OB association  and hosts two transiting exoplanets: a 10~R$_{\earth}$ planet in a 6.96~d orbit and a 7.9~R$_{\earth}$ planet in a 14.3~d orbit \citep{rizzutoetal20-1,barber24-1}. Rizzuto et al.\ determined a 17~Myr age and $T_{eff} = 5675$~K for the host star, making it the youngest transiting giant planet system and an excellent testbed for the evolution of gaseous planets in a very young solar analogue system. \citet{thao24-1} obtained near-infrared (NIR) transmission spectroscopy of HIP~67522b  with JWST NIRspec. They determined that while the radius of the planet is Jupiter-sized, the depth of its absorption features were much larger than would be expected for a mature Jupiter. They determine a mass of $13.8\pm0.1 M_{\earth}$ and thus a density of $<0.1$~g~cm$^{3}$, making HIP~67522b one of the lowest density planets known. 

In this paper, we present the XUV spectrum and variability of \targ, evaluating the stellar environment in a young solar analogue at the epoch of most significant planetary mass loss. The observations and data reduction are presented in Section~\ref{sec:obs}, Section~\ref{sec:analysis} covers the data analysis, and Sections~\ref{sec:discussion} and~\ref{sec:conclusion} present our discussions and conclusions.

\section{Observations and Data Reduction} \label{sec:obs}

\subsection{Ultraviolet}  

We observed \targ\ with HST STIS \citep{woodgateetal98-1} as part of GO program \#16701; overall results from that program are presented in Wilson et al. (2025, submitted) and will be available at the Mikulski Archive for Space Telescopes (MAST): doi:10.17909/T9DG6F. The observations were performed in 2022 on September 03 and 05 using the G140L, G230L, and G430L gratings with a $52\times0.2$~arcsec slit aperture to obtain far-ultraviolet (FUV), near-ultraviolet (NUV), and optical spectroscopy at a resolving power of R=1000 in the FUV and R=500 in the NUV and visible. The UV observations were executed in the time-tagged observing mode, providing the arrival time of each photon to within 125~$\mu$s. The observations are summarized in Table~\ref{tab_obs}. The UV and X-ray data were first presented in their current form in \citet{thao24-1}, but we re-summarize the data reduction steps here for reference.

\begin{deluxetable}{ccccccc}
\tablecaption{Observation Summary\label{tab_obs}}
\tabletypesize{\small}
\tablecolumns{4}
\tablehead{\colhead{Observatory} & \colhead{Instrument Setting}  & \colhead{Date} & \colhead{T$_{exp}$ (sec)}  }
\startdata
HST & STIS G140L & 2022-09-03 &  9606 \\
HST & STIS G140L & 2022-09-05 & 4587 \\
HST & STIS G230L & 2022-09-05 & 2103 \\
HST & STIS G430L & 2022-09-05 & 10 \\
HST\tablenotemark{a} & COS G130M & 2022-07-12 & 5564 \\
Chandra\tablenotemark{a} & ACIS-S4 & 2021-02-08 & 2080 \\
\hline
\enddata
\tablenotetext{a}{Archival data.}
\end{deluxetable}

\begin{figure}
    \centering
    \includegraphics[width=0.9\textwidth]{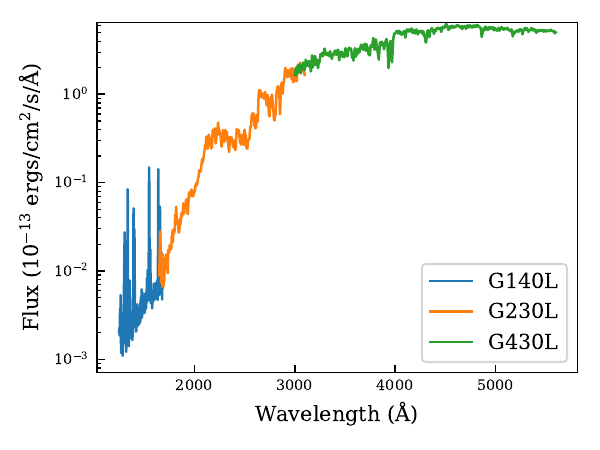}
\caption{The UV-optical spectrum of \targ, obtained with HST STIS. The spectra have been dereddened assuming A$_{V} = 0.12\pm0.06$~mag \citep{rizzutoetal20-1}. }
    \label{fig:allspec}
\end{figure}

\begin{figure}
    \centering
    \includegraphics[width=0.9\textwidth]{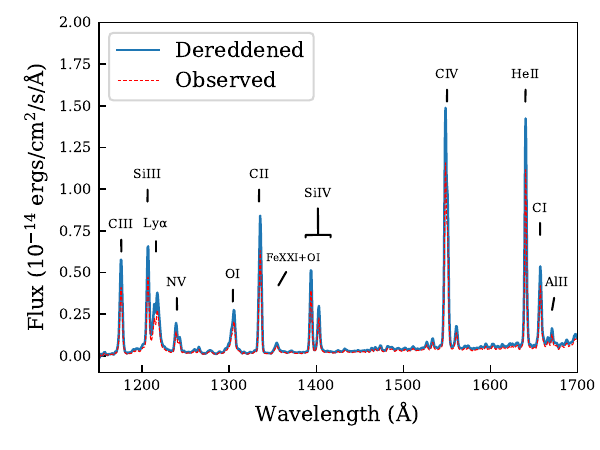}
\caption{The STIS FUV spectrum of \targ. The observed and dereddened spectra are shown in red and blue, respectively. Prominent FUV emission lines are labelled. }
    \label{fig:fuvspec}
\end{figure}

After examining the individual spectra and comparing them to separate COS observations of the same target (see below), we found that the Calstis data reduction pipeline was inducing errors of $\simeq$100~km~s$^{-1}$ in the individual G140L spectra. To correct, we re-extracted each spectrum after supplying the correct SHIFTA1 keyword and disabling the WAVECORR keyword. We coadded the G140L exposures weighted by flux uncertainty (to deweight low signal to noise data near the detector edges) and then spliced spectra from all three grating settings together. Where two spectra overlapped, we used the data from the blueward grating. The STIS spectra are presented in Figures~\ref{fig:allspec} and \ref{fig:fuvspec}.

\targ\ was also observed with COS on 2022 July 12 under GO program \#16901 and published in \citet{maggio24-1}. We retrieved the COS spectrum --- only one exposure in the visit had usable data --- to compare the average spectra from both STIS and COS observations, taken two months apart. We dereddened both spectra using the python dust\_extinction package \citep{gordon23-1,Gordon2024},  adopting A$_{V} = 0.12$~mag \citep{rizzutoetal20-1} and a standard extinction relation (R$_{V} = 3.1$). In Table~\ref{tab:lines} we present the dereddened line fluxes for both the STIS and COS observations. The two spectra are similar, with modestly higher ($\simeq$15--25\%) line fluxes in the STIS data.

\begin{deluxetable}{ccccccc}
\tablecaption{UV Dereddened Line Fluxes\label{tab:lines}}
\tabletypesize{\small}
\tablecolumns{5}
\tablehead{ & & \colhead{STIS} & \colhead{COS} \\
\colhead{Transition} & \colhead{Wavelength}  & \colhead{Flux (10$^{-14}$)} &   \colhead{Flux (10$^{-14}$)} \\
& (\AA) & (ergs~cm$^{-2}$~s$^{-1}$) & (ergs~cm$^{-2}$~s$^{-1}$)}
\startdata
\ion{C}{3}  & 1176\tablenotemark{a} & 2.00$\pm$0.05 & 1.76$\pm$0.03 \\
\ion{Si}{3} & 1206.5 & 2.03$\pm$0.05 & 1.62$\pm$0.04 \\
\ion{N}{5} & 1238.8 & 0.54$\pm$0.02&  \\
\ion{N}{5} & 1242.8 & 0.27$\pm$0.02 & 0.21$\pm$0.02 \\
\ion{O}{1} & 1304\tablenotemark{b} & 1.61$\pm$0.08 & \nodata\tablenotemark{c}\\
\ion{C}{2} & 1334\tablenotemark{d} & 2.77$\pm$0.06 & 2.37$\pm$0.06 \\
\ion{Cl}{1} & 1351.7 & \nodata & 0.06$\pm$0.01\\ 
\ion{Fe}{21} & 1354.1 & 0.29$\pm$0.1\tablenotemark{e} & 0.14$\pm$0.02 \\
\ion{O}{1} & 1355.6 & \nodata\tablenotemark{e} & 0.08$\pm$0.02 \\ 
\ion{Si}{4} & 1393.8 & 1.42$\pm$0.05 & 1.22$\pm$0.04 \\
\ion{Si}{4}  & 1402.8 & 0.84$\pm$0.03 & 0.61$\pm$0.03\\
\ion{C}{4} & 1548.2 & 4.5$\pm$0.4 \\
\ion{C}{4} & 1550.8 & 1.4$\pm$0.3 \\
\ion{He}{2} & 1640.4 & 3.5$\pm$0.1 \\
\ion{C}{1} & 1657.0 & 0.13$\pm$0.03 \\
\ion{Al}{2} & 1670.8 &  0.24$\pm$0.02 \\
\hline
\enddata
\tablenotetext{a}{Multiplet. Fit with one Gaussian for STIS spectrum and six for COS, shown here as a summed line flux.}
\tablenotetext{b}{Multiplet. Fit with three Gaussians, shown here as a summed line flux.}
\tablenotetext{c}{Airglow contaminated.}
\tablenotetext{d}{Multiplet. Fit with one Gaussian for STIS spectrum and three for COS, shown here as a summed line flux. }
\tablenotetext{e}{In the STIS spectrum \ion{Fe}{21} and \ion{O}{1} are blended. The combined flux is given. }
\tablecomments{Dereddened with R$_{V}$ = 3.1 and A$_{V}$ = 0.12. COS data only comprise G130M observations so longer wavelength data is unavailable. }
\end{deluxetable}

\subsection{X-ray}

We used archival Chandra data to calculate the X-ray emission from \targ. It was observed in 2021 for $\simeq$2~ksec (OBSID 24675 PI: G. Garmire). 
The count rate was $0.093\pm0.007$~cnts~s$^{-1}$ with no detectable sign of variability. The 0.3--10~keV flux was observed to be $1.20\pm0.09 \times 10^{-12}$~ergs~cm$^{-2}$~s$^{-1}$, or an X-ray luminosity of $L_{X} = 2.2\pm0.2 \times 10^{30}$~ergs~s$^{-1}$ ($\log L_{X} = 30.35$). We fit a single temperature spectrum to the data using XSPEC (V. 12.13.0). The fit assumed an interstellar absorption value of $\log{N_{H}}=2.1\pm1.1\times10^{20}$~ergs~cm$^{-2}$~s$^{-1}$~\AA, consistent with the optical extinction ($A_{V}$) value from \citet{rizzutoetal20-1}. The derived X-ray parameters were insensitive to the exact value of $N_{H}$ within this range.
The XSPEC fits estimated a coronal temperature of $1.16\pm0.08$~keV ($\sim13$~MK).  As noted in \citet{thao24-1}, the log ratio of the observed X-ray luminosity in the ROSAT band (0.1--2.4~keV) to the bolometric luminosity is -3.53, a ratio that corresponds to the ``saturated'' coronal activity expected for such a young star \citep{pizzolato03-1,wright11-1}.

\targ\ was also observed by XMM, with the results presented in \citet{maggio24-1}. They obtained 70~ksec of data using EPIC (as well as the OM/UVM2 to obtain NUV light curves). The longer observing sequence showed significant variability, including two large X-ray flares. Their quiescent spectral fits yield luminosities of $\log L_{X} = 30.46-30.55$, 30\% -- 60\% higher than what was seen in the Chandra observations. The Maggio et al.\ mean plasma temperature for the long quiescent phase is 15.4~MK, slightly hotter than the value we found, which would be in agreement with the larger L$_{X}$ seen during their observation.

\section{Analysis} \label{sec:analysis}

 \subsection{Time Variability}

\begin{figure}
    \centering
    \epsscale{1.15}
    \plottwo{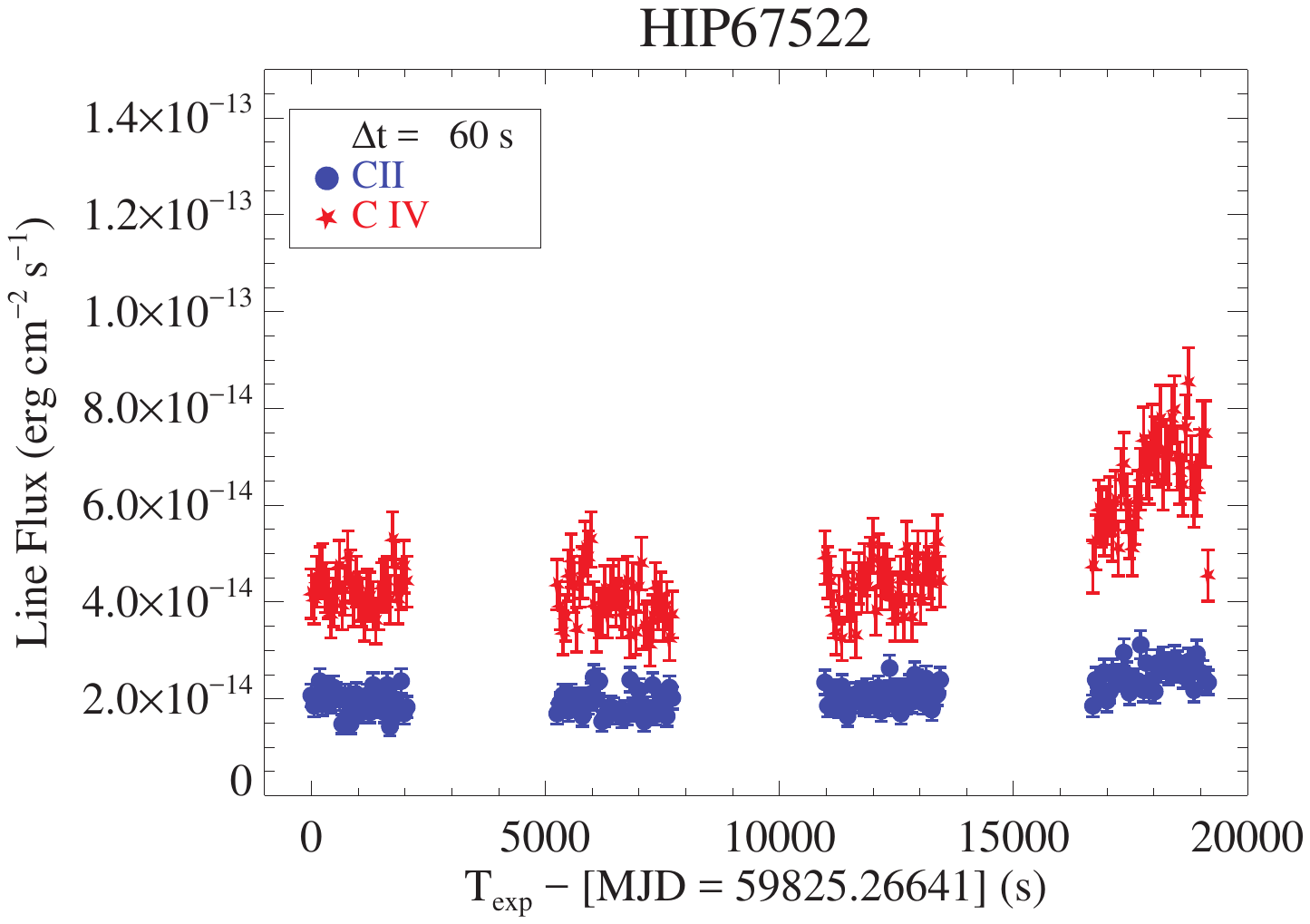}{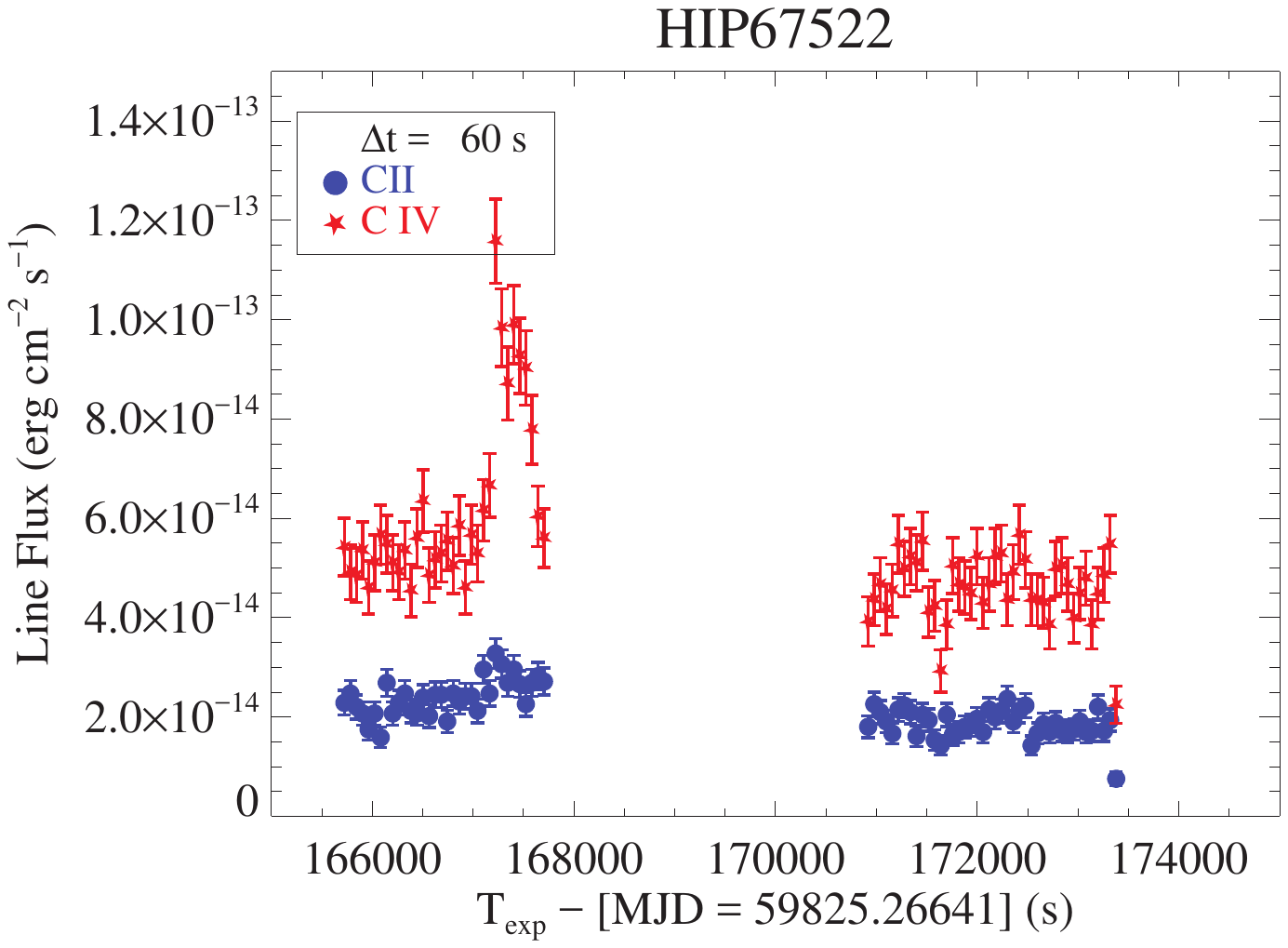}
\caption{Line flux measurements for \ion{C}{2} and \ion{C}{4} are shown, binned into 60~sec increments. }
    \label{fig:varib2}
\end{figure}

We averaged the time-tagged spectra into 60~sec bins to investigate variability during the STIS G140L observations. 
The UV spectrum exhibited continuous low-level variability ($\simeq$15\% about the mean) as well as two small flares (near 18~ksec and 170~ksec). In Figure~\ref{fig:varib2} we compare the line flux changes in two of the brightest lines, \ion{C}{2} and  \ion{C}{4}, which probe two different formation temperatures ($\log T = 4.5$~K and $\log T = 5$~K, respectively). The two flares are morphologically different: while the second flare shows a classic impulsive rise, the first has a more gradual increase over several hundred seconds. In both cases, the flare is strongest in \ion{C}{4}, which more than doubles in strength, while \ion{C}{2} varied by $\simeq$25\%.



We calculated the equivalent duration of the two flares. The equivalent duration, $\delta$, is defined as the time required for the quiescent emission from the star to equal the energy released in the flare. We calculated $\delta$ for the broadband FUV, 1233--1688~\AA\ (excluding Ly$\alpha$ to remove the effects of airglow) and in \ion{C}{4} and tabulated the results in Table~\ref{tab_flares}. In the construction of the DEM  we include all of the UV data, flare and quiescent, as the time-averaged behavior of the star will incorporate this type of activity. 

\begin{deluxetable}{cccc}
\tablecaption{Flare Energies and Equivalent Durations\label{tab_flares}}
\tablecolumns{4}
\tablehead{\colhead{Flare} & Band & \colhead{$\log_{10}E$}  & \colhead{$\delta$}  \\
 \colhead{\#} & &  \colhead{(ergs)} & \colhead{(sec)} }
\startdata
Flare 1 & FUV & 32.68 & 937.6 \\
& CIV & 31.93 & 1239.7 \\ 
Flare 2 & FUV & 32.17 & 291 \\
& CIV & 31.58 & 561.4 \\ 
\enddata

\end{deluxetable}

\subsection{Ly~$\alpha$ Reconstruction}\label{sec:lya}
\subsubsection{Reconstruction Results Using the Standard Fitting Process }

\ion{H}{1} 1215.67 \AA\ \lya\ is the dominant stellar emission line in the UV. Because the interstellar medium absorbs the \ion{H}{1} emission, the intrinsic flux emitted by the star and seen by its planets must be reconstructed. To do so, we fit a model comprising the intrinsic stellar \lya\ flux and the interstellar absorption component convolved with the instrument line function to the STIS G140L spectra. The model and fitting process are described in detail in \cite{youngbloodetal21-1,youngbloodetal22-1}.   They also discuss the uncertainties on the reconstructed \lya\ profile as a function of the spectral resolution, signal to noise ratio (SNR), and intrinsic width of the stellar line.  In short, we use a likelihood-based Bayesian calculation and the \texttt{emcee} Markov Chain Monte Carlo (MCMC) implementation \citep{foreman-mackeyetal13-1} to fit Equation 3 of \cite{youngbloodetal22-1} to the data. Because the wings of \lya\ and \ion{Si}{3} are blended in our spectra, we also include a Gaussian emission line convolved with the instrument line function in the model. We assume uniform priors on all fitted parameters unless stated otherwise. 

Figure \ref{fig:lya1} shows two \lya\ reconstruction fits. The corner plots for the fits are shown in the Appendix in Figures~\ref{fig:lya1corner} --~\ref{fig:lya2corner}. The left panel gives the results of the standard fitting process. There are no constraints on the radial velocity, amplitude, line widths, self-reversal parameter, interstellar column density or the radial velocity of the ISM (defined as an offset from the stellar radial velocity), except that the latter is set not to exceed $\pm50$~km~s$^{-1}$. The Doppler ``b" value and D/H ratio are fixed at typical values for the local ISM, (11.5 km~s$^{-1}$  and $1.5\times10^{-5}$, respectively).

\begin{table}
    \centering
     \caption{\lya\ Intrinsic Line Flux Estimates}
    \label{tab:lyatab}
    \begin{tabular}{lcccc}
\hline \hline
Flux Estimate Source & $f_{\mathrm{Ly\alpha}}$  & 
$\log$\,N(\ion{H}{1})  & $\chi^{2}_{\nu}$ & Ref. \\
 & (10$^{-14}$ erg s$^{-1}$ cm$^{-2}$) & (cm$^{-2}$) \\
\hline 
Reconstruction, free fit & $4.2^{+1.0}_{-0.8}$ & $18.80^{+0.27}_{-0.63}$ & 3.9 & This work  \\ 
Reconstruction, constrained fit 1 & $7.2^{+0.4}_{-0.5}$ & 19.36\tablenotemark{b} & 7.1 & This work \\
Reconstruction, constrained fit 2\tablenotemark{a} & $490^{+600}_{-25-}$ & 20.0\tablenotemark{b} & 9.8 & This work \\
Reconstruction, four IS absorbers & $7.4^{+0.6}_{-0.5}$ & 19.47$\pm$0.03\tablenotemark{c} & 6.4 & This work \\
\ion{C}{4} scaling & 57.8 & & & Linsky et al.\ (2013) \\
X-ray scaling & 18.7 & &  & Linsky et al.\ (2013) \\
X-ray young star & 31.0 & &  & Linsky et al.\ (2020)\\
Stellar rotation scaling & 58.6\tablenotemark{d} & & & Wood et al.\ (2005) \\
\hline
   \end{tabular}
\tablenotetext{a}{Fit did not converge. The values shown here are those recorded after 100,000 iteration steps.}
\tablenotetext{b}{Value fixed in fit.}
\tablenotetext{c}{Each absorber was constrained to have a value of $\log N_{HI} \geq 18.78$ (rounded to 18.8 in the fit) so that the total absorption column along the line of sight was $\log N_{HI} \geq 19.36$. The median $\log N_{HI}$ for each of the four components in the best fit were 18.84, 18.85, 18.85, and 18.85. The low uncertainty on the interstellar column density is due to the fit piling up near the lower limit for each absorber.}
\tablenotetext{d}{This value was used in the construction of the SED presented in \citet{thao24-1}.}
\tablecomments{ $f_{\mathrm{Ly\alpha}}$ is the estimated of the observed flux of \lya\ at Earth in the absence of interstellar absorption.} 
\end{table}

\begin{figure}
    \centering
    \includegraphics[width=0.9\textwidth]{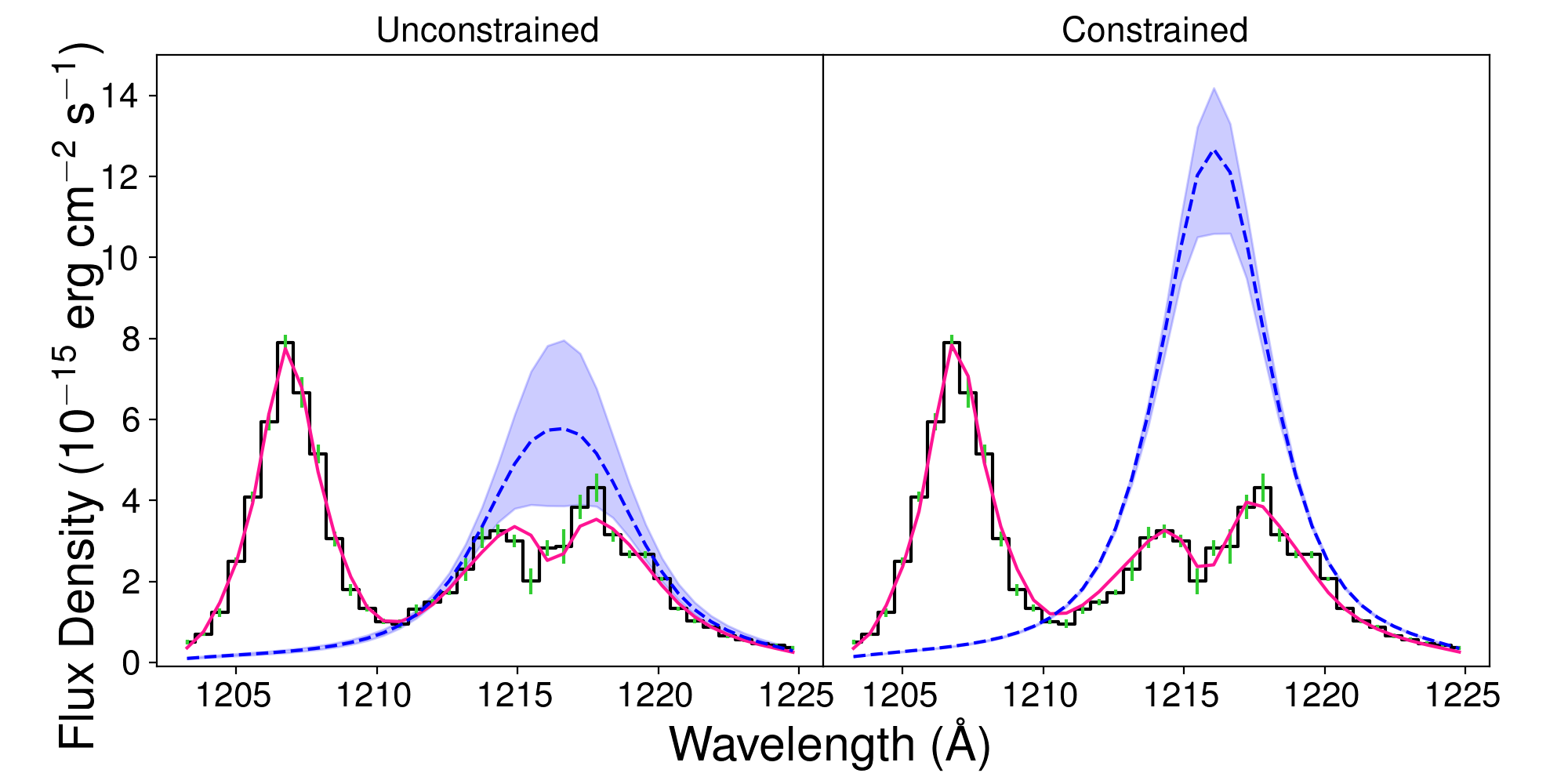}
    \caption{Ly$\alpha$ reconstructions. In both panels, the STIS data are shown in black and the model fits (intrinsic Ly$\alpha$ profile folded through the ISM) in red. The dotted blue lines show the reconstructed, intrinsic stellar \lya\ emission with $1\sigma$ uncertainties in shadow. The left panel shows the unconstrained parameter fitting. The reconstructed (unattenuated) Ly$\alpha$ flux is $4.2^{+1.0}_{-0.8} \times 10^{-14}$~erg~cm$^{-2}$~s$^{-1}$ for a derived ISM absorption of $\log(N_{HI}) = 18.80^{+0.27}_{-0.63}$. The right panel shows the fit when the interstellar column density is fixed to $\log(N_{HI}) = 19.36$. The reconstructed (unattenuated) Ly$\alpha$ flux is $7.19^{+0.42}_{-0.53} \times 10^{-14}$~erg~cm$^{-2}$~s$^{-1}$. 
    }
    \label{fig:lya1}
\end{figure}
 
 While the statistical quality of the standard fit is acceptable ($\chi^{2} = 108$ with 28 degrees of freedom, or $\chi^{2}_{\nu} = 3.9$) , the low amplitude of the Ly$\alpha$ flux ($4.2^{+1.0}_{-0.84} \times 10^{-14}$~erg~cm$^{-2}$~s$^{-1}$) is unphysical relative to the strength of the adjacent \ion{Si}{3} 1206 line, given the relative abundances of the two elements in the Sun \citep{lodders03-1}. \targ\ would require highly anomalous silicon to hydrogen abundance ratios to generate such a pronounced deviation, while \citet{rizzutoetal20-1} note that the young regions around Sco-Cen have metallicities consistent with solar. (For comparison, see Figure~\ref{fig:tomplots} below, where the spectrum of \targ\ is compared to that of the 100~Myr G5V star EK~Dra and the 45~Myr G6V star DS~Tuc. Even when attenuated by the ISM, \lya\ is significantly stronger than \ion{Si}{3} in young, solar type stars.)


 To test if a degeneracy in the fitting process was driving the low value of the reconstructed flux, we re-ran the fit with two different fixed values of the IS absorption. First, we fixed the interstellar column density to the best-fit value found by \citet{maggio24-1} in their model fit to the X-ray spectrum of \targ. Their column density, $\log(N_{HI}) = 19.36$, is 3.6$\times$ the $N_{H}$ value returned by the unconstrained reconstruction. We also forced the radial velocities of \ion{Si}{3} and \lya\ to be identical. The ISM radial velocity (defined as an offset from the stellar radial velocity) was constrained by a Gaussian prior centered at -30$\pm$5 km s$^{-1}$. This prior is based on the archival COS spectrum of C II (1334 \AA), where the ISM absorption is approximately -30 km s$^{-1}$ Doppler shifted from the stellar line. 
	
 The reconstructed \lya\ flux in the new fit (``Constrained fit 1" in Table~\ref{tab:lyatab}) increased by a factor of 1.7, to $7.2^{+0.4}_{-0.5} \times 10^{-14}$~erg~cm$^{-2}$~s$^{-1}$, but there remain several indications that the fit is unreliable. First, the statistical quality of the fit degraded (from $\chi^{2}_{\nu}$ = 3.9 to 7.1,  with 28 and 30 degrees of freedom, respectively). Second, in order to increase the \lya\ intrinsic flux, several other free parameters had to be pushed to their numerical limits. Finally, the wings of the reconstructed line do not match the observed data at high velocity, indicating that either that the fit is spurious and/or we have no observational information to constrain the intrinsic line flux. 

 The $\log(N_{HI}) = 19.36$ value found by Maggio et al.\ is lower than the value of $\log(N_{HI}) = 20.04$ found by scaling the neutral hydrogen column density from the visible dust extinction value of $A_{V} = 0.12$ \citep{thao24-1}. X-ray absorption traces metals rather than neutral hydrogen, and X-ray spectral fits are not sensitive to the exact value of $\log(N_{HI})$ at these column densities. On the other hand, the optical extinction is due to dust rather than gas, so it also does not directly trace the neutral hydrogen columnn density. Accordingly, we also ran a reconstruction wherein we fixed $\log(N_{HI}) = 20.0$ to span the range of absorption values indicated by the disparate estimation methods. That fit is presented in Table~\ref{tab:lyatab} as ``Reconstruction, constrained fit 2." It did not converge on a solution, however. After 100,000 iteration steps, the intrinsic \lya\ flux was at  $4.9^{+6.0}_{-2.5} \times 10^{-12}$~erg~cm$^{-2}$~s$^{-1}$ with a reduced $\chi^{2}_{\nu}$ = 9.7, indicating an even greater deviation from the observed data as the IS absorption was increased.

\subsubsection{Interstellar Absorption Measured from Nearby Lines of Sight to HIP~67522 } \label{sec:ism}

One potential cause of anomalous fits above may be due to excess, high velocity absorption that is not captured in the default models.  \targ\ is more distant and has a higher \ion{H}{1} column density than targets previously observed at low spectral resolution, all of which were within 20~pc, so our analysis is pushing to stronger attenuation than investigated by  \citet{youngbloodetal21-1}.  In order to better understand the likely interstellar (IS) absorption towards \targ\, we retrieved high resolution HST near-ultraviolet (NUV) spectra for four early-type stars (three rapidly-rotating A stars and 3~Cen~A, a B5 III star) within 15 degrees of \targ. These STIS E230H spectra contain interstellar absorption lines of \ion{Mg}{2} and \ion{Fe}{2} that provide information on the IS clouds that are likely to absorb the stellar \lya\ emission. Table~\ref{tab_ism} lists the relevant properties of the AB stars and \targ. The stars have distances between 71 and 152 pc, which brackets the 125 pc distance of \targ. 

\begin{deluxetable}{cccccccc}
\tablecaption{ISM Along the Line of Sight to HIP 67522\label{tab_ism}}
\tabletypesize{\small}
\tablecolumns{8}
\tablehead{\colhead{Star} & \colhead{Spectral}  & \colhead{T$_{eff}$} & \colhead{Distance}\tablenotemark{a} &
\colhead{Age} & \colhead{Association}\tablenotemark{b} & \colhead{r}\tablenotemark{c} & \colhead{V$_{ISM}$} \\
    & \colhead{Type} & \colhead{(K)} & \colhead{(pc)} & \colhead{(Myr)}&  & \colhead{(Degree)}&
 \colhead{(km~s$^{-1}$)}}
\startdata
HIP 67522  & G0 V    &  5675 &  124.7$\pm$0.3 &  17 &  UCL/LCC  &  0.0 & \nodata \\
HD109573  & A0 V    &  9670  &    70.8$\pm$0.2 &  10 & TWA                & 14.1 & -14/-5 \\
3 Cen A      & B5 IIIp  &17500 &  100.9$\pm$1.7 &      &  UCL/LCC        &  7.8 & -21/-8/+5/+14\\
HD110058  & A0 V    &  7990 &  130.1$\pm$0.5 &  17 & UCL/LCC        & 14.9 & -15/+0+5/+12\\
HD131488  & A1 V    &  8700 &  152.2$\pm$0.8 & 16 & UCL/LCC         & 12.3 & -23/-15/-5/+5\\
\hline
\enddata
\tablenotetext{a}{GAIA DR3 \citep{vallenari23-1}}
\tablenotetext{b}{\citet{luhman22-1}. UCL/LCC: Upper Centaurus-Lupus/Lower Centaurus-Crux. TWA: TW Hydrae association.}
\tablenotetext{c}{Distance on the sky from HIP~67522 in degrees.}

\end{deluxetable}

\begin{figure}
    \centering
    \includegraphics[width=0.9\textwidth]{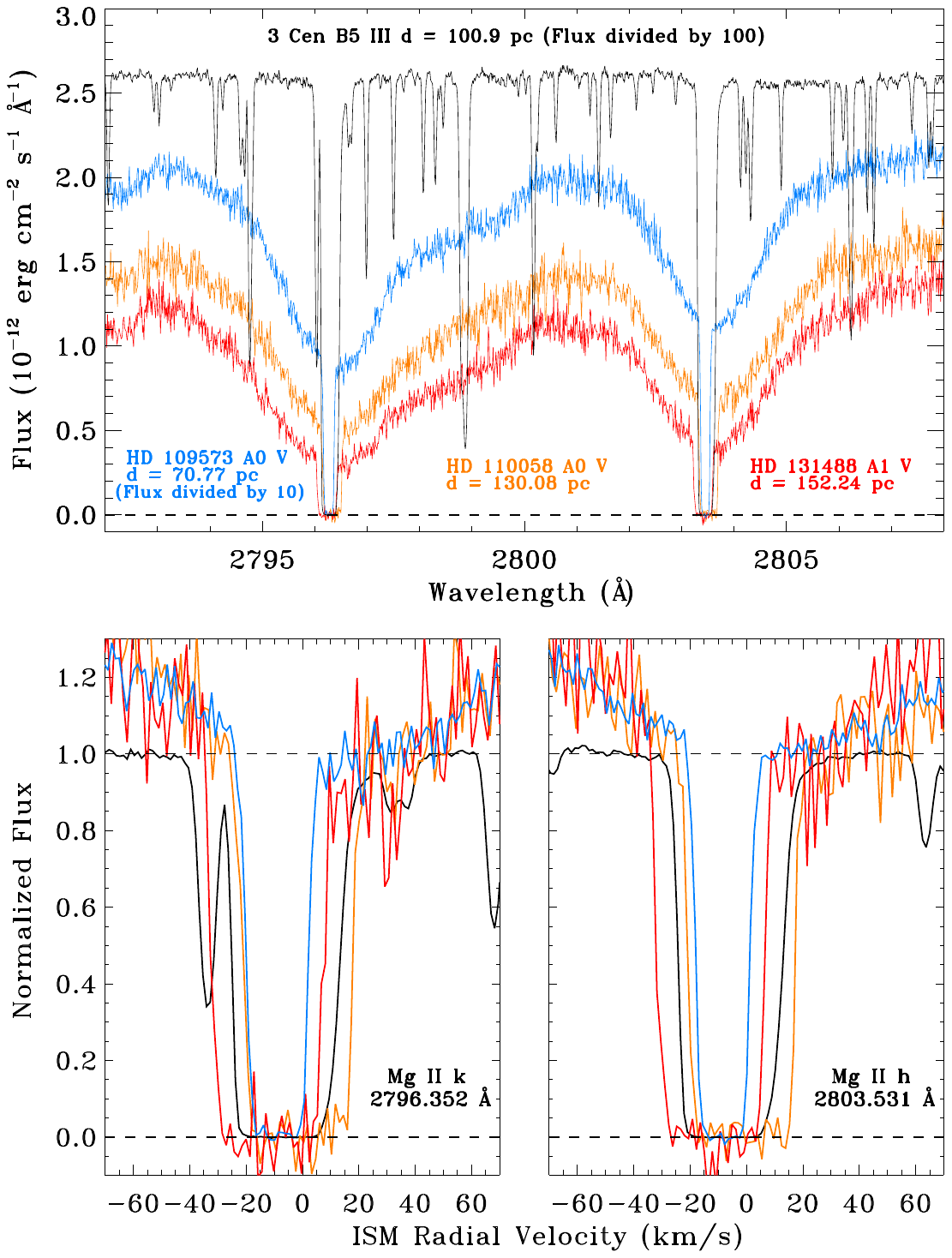}
    \caption{Interstellar absorption along lines of sight within 15$^{\circ}$ of Sco-Cen as probed by 3~Cen, HD1095793, HD110058, and HD131488. The top panel shows the NUV spectrum around \ion{Mg}{2}. The bottom panel shows a close-in plot of \ion{Mg}{2} plotted in velocity space. None of the nearby sight lines show interstellar absorption beyond $\simeq30$~km~s$^{-1}$.}
    \label{fig:ism}
\end{figure}

\begin{figure}
    \centering
    \includegraphics[width=0.7\textwidth,angle=90]{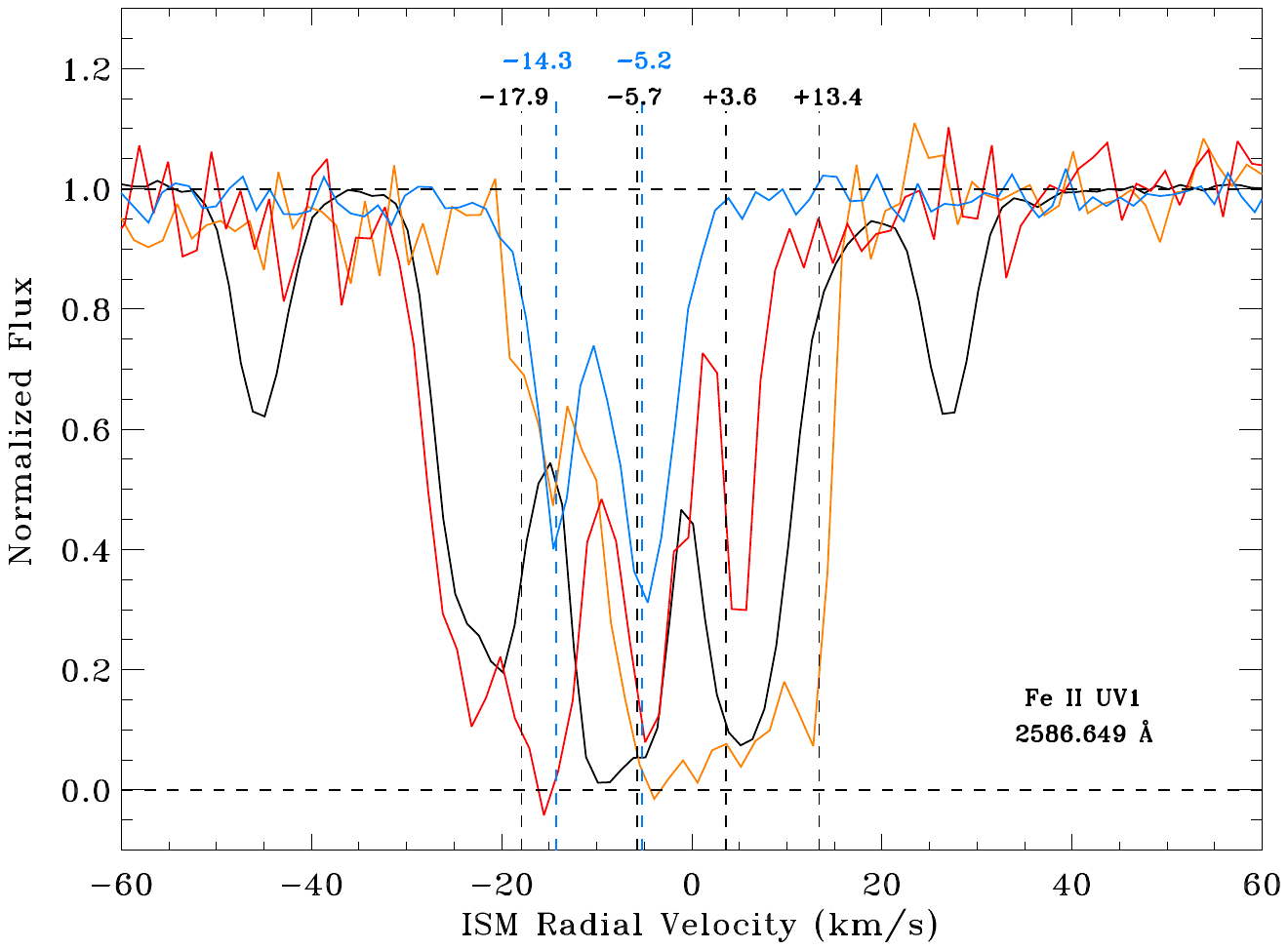}
    \caption{Interstellar absorption along lines of sight within 15$^{\circ}$ of Sco-Cen as probed by 3~Cen (black), HD1095793 (blue), HD110058 (orange), and HD131488 (red). This figure shows \ion{Fe}{2} 2586 absorption in velocity space. Prominent absorption components are labeled: the velocities in blue are previously-identified absorbers in nearby lines of sight, while the black lines are the average velocities from the four reference stars used in the \lya\ reconstruction. }
    \label{fig:ismfe2}
\end{figure}

Figures~\ref{fig:ism} and~\ref{fig:ismfe2} show their NUV spectra centered on the \ion{Mg}{2} 2800 doublet and the \ion{Fe}{2} 2586 line, respectively.
The IS absorption for all the stars is complex, with four separate IS clouds present at velocities between  $-23$ and +14 km s$^{-1}$. There are two components (at -5.2 and -14.3~km~s$^{-1}$) present in the spectrum of the closest star, HD~109573 (71~pc); these absorbers were also identified in Local Interstellar Medium study of \citet{nisak25-1}. 
As distances increase above 100~pc, at least two other components appear around -20 and  +5~km~s$^{-1}$. The latter is seen in all four stars, though the -20~km~s$^{-1}$ absorber is not observed in HD110058. 3~Cen~A is nearest on the sky to \targ\ but closer (100~pc vs.\ 125~pc), suggesting that all four of the observed IS absorption components seen in the former will also be present in the latter. We undertook simultaneous fitting of the \ion{Mg}{2} and \ion{Fe}{2} absorption features for the four reference stars and identified four absorbers likely to be present along the line of sight to \targ, at -17.9~km~s$^{-1}$, -5.7~km~s$^{-1}$, +3.6~km~s$^{-1}$, and +13.4~km~s$^{-1}$. 

\subsubsection{Reconstruction Results Using a Targeted IS Absorption Profile}

Armed with a better picture of the likely composition of IS clouds along the line of sight to \targ, we undertook a final round of \lya\ reconstruction with four IS absorbers explicitly included in the fit. We ran fits both with fixed values of the IS velocities and where they were allowed to vary with a Gaussian prior centered on the starting velocities. We forced the Doppler ``b" value for each cloud to be 11.5~km~s$^{-1}$. The neutral hydrogen column density was allowed to vary, although we set the minimum value for each absorber to  $\log N(HI) > 18.8$  such that the total IS absorption was consistent with the X-ray estimate from \citet{maggio24-1} or higher. The new fits  returned a reconstructed \lya\ emission flux of $7.4^{+0.6}_{-0.5} \times 10^{-14}$~erg~cm$^{-2}$~s$^{-1}$ with a profile similar to the right hand panel in Figure~\ref{fig:lya1}. This result is consistent with the values from the ``Constrained fit 1"  and remains inconsistent with our expectations of the \lya\ emission from a 17~Myr star, a point we expand upon further in the next section.

\subsubsection{\lya\ Fluxes from Scaling Relations} \label{sec_scaling}
 
To get a better estimate of the expected \lya\ emission from \targ, we compare our fits to estimates of the \lya\ flux based on several scaling relations. In Table~\ref{tab:lyatab}, we compare the reconstructed line fluxes to values predicted from other measurements. The table presents the integrated line fluxes that would be observed at Earth in the absence of interstellar absorption. We used the scaling relations from  \citet[][their Table 2]{linsky13-1}  for the \ion{C}{4} FUV line and X-ray flux values. We also estimated how much stronger \lya\ would be given the young age of \targ\ based on \citet{linskyetal20-1}, who showed that \lya\ could be another $\simeq$60\% brighter for the youngest solar-like stars. Finally, we used the scaling relation between \lya\ surface flux and the stellar rotation period for G and F stars from \citet{woodetal05-1} for our ``Stellar rotation scaling" estimate, using a rotation period of 1.418~d \citep{rizzutoetal20-1}. A comparison of the reconstructed \lya\ intrinsic fluxes to those estimated from empirical scaling relations shows that the reconstructed values are $\simeq4$--30 times lower than would be expected for a young G star, further demonstrating the problems with the reconstruction process for these observations. For the SED presented in \citet{thao24-1}, we used the stellar rotation scaling prediction for the \lya\ flux.

\section{Discussion} \label{sec:discussion}

\subsection{Intrinsic \lya\ Emission in \targ\ and Likely Local Absorption }


The evidence from empirical relationships between X-ray and UV line emission and \lya\ line strength as well as the problems with our \lya\ fits indicate that it is not possible to reconstruct the intrinsic line profile with any confidence from our data,  an issue that has not previously been experienced with the \lya\ reconstruction technique \citep{woodetal05-1,franceetal13-1,youngbloodetal22-1,bourrieretal17-1,bourrieretal18-1}. To understand why the reconstruction technique was unable to obtain an accurate determination of the unabsorbed \lya\ flux, we first examine potential issues with the method itself. One limitation of the analysis here may be the low spectral resolution of the data. \citet{youngbloodetal21-1} discusses the application of the \lya\ reconstruction technique to low resolution spectra. They found that when they degraded high resolution spectra (from STIS E140M, R = 45,800) to the resolution of the G140L data (R=1000), the
best fit fluxes were not replicated at lower resolution, but overlapped within at least the 95\% level (roughly 2$\sigma$), albeit with degraded precision. However, the worst fits were for stars with intrinsically narrow \lya, such that there was insufficient information to resolve the degeneracy between the intrinsic line strength and the interstellar absorption, which is not the case for this target.

\citet{wilson22-1} further took advantage of a M dwarf-white dwarf binary system to further test the accuracy of \lya\ reconstruction by comparing the predicted line profile to the actual emission seen when \lya\ Doppler-shifted out of the interstellar absorption band. They found that a degeneracy between the intrinsic Ly$\alpha$ and the ISM profiles could cause the reconstructed flux to be off by up to a factor of two at the lowest line velocity. Both Youngblood et al.\ and Wilson et al.\  identify issues with the reconstruction when the line core is completely occulted. However, their errors are at the level of a factor of two, while we find reconstructed fluxes greater than an order of magnitude below scaling relation predictions, indicating a deeper cause. Our attempts to improve the fits using a more realistic assessment of the IS absorption profile failed to resolve this discrepancy.

\begin{figure}
    \centering
    \epsscale{1.3}
    \plottwo{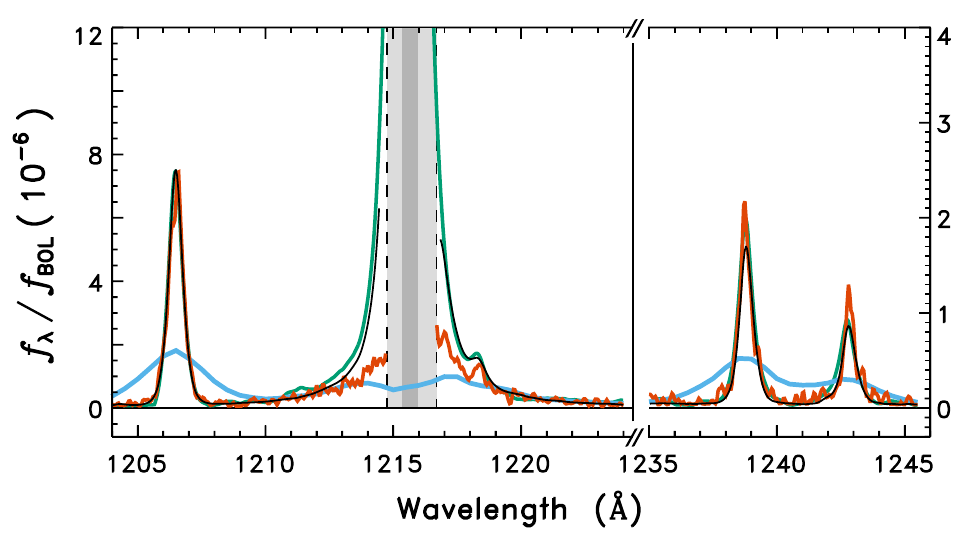}{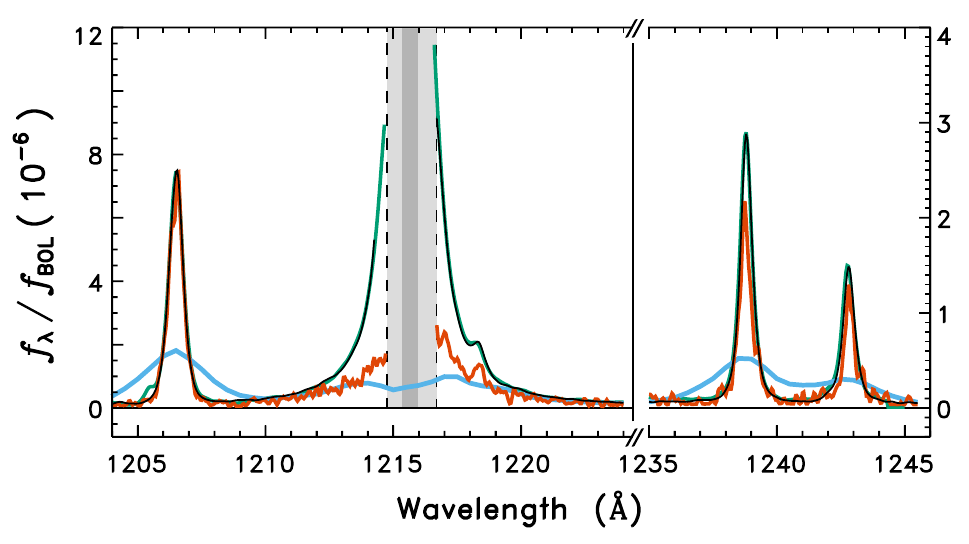}
    \caption{A comparison of the COS and STIS spectra of \targ\ with those of EK~Dra (top) and DS~Tuc (bottom). The left inset in each panel shows \ion{Si}{3} 1206 and \lya, while the right has the \ion{N}{5} doublet; note the y-axis scale change to accommodate the fainter N V emissions. For \targ, In both panels, red curve is the COS G130M spectrum and the blue curve is the STIS G140L spectrum. The spectra have been normalized by their bolometric fluxes, $f_{BOL}$.}
    In the top panel, the COS G130M spectrum of EK~Dra is in black and the 
    STIS E140M spectrum in green. 
    In the bottom panel, the COS G130M spectrum of DS~Tuc is in black and the STIS G140M in green.
    The vertical dashed lines show the approximate wavelength range normally affected by geocoronal \lya\ for COS G130M; the inner darker shaded band is where the IS \lya\ absorption occurs (based on STIS  medium-res E140M of EK Dra, d~34 pc). 
    In both plots, the wavelength scales were registered by adjusting the flux-weighted wavelength of \ion{Si}{3} to the laboratory value 1206.500~\AA. The y-axis is the  ISM-corrected bolometrically normalized flux for \targ, 
    while the other star has been smoothed in wavelength, and scaled in $f/f_{BOL}$, to match the apparent line shape of \ion{Si}{3} \targ. The smoothing utilized a pseudo-Gaussian profile: normal exponent alpha=2 replaced by alpha= 1.25  yielding a sharper core plus broader wings than normal Gaussian (based on empirical experimentation).  
    \label{fig:tomplots}
\end{figure}

With the above in mind, we turn our attention to the \targ\ system itself. In Figure~\ref{fig:tomplots}, we compare the FUV spectrum of \targ\ to that of two other young, solar-type stars, EK~Dra and DS~Tuc. The three stellar spectra are normalized by their bolometric fluxes; these were computed using the methodology presented in \citet{ayres22-1}, resulting in $f_{BOL}$ = 0.366, 2.31, and 1.11 ($\times 10^{-8}$~ergs~cm$^{-2}$~s$^{-1}$) for \targ, EK~Dra, and DS~Tuc, respectively.  We scaled the comparison stars' spectra to fit the shape and intensity of the \ion{Si}{3} 1206 line in \targ. \ion{Si}{3} 1206 is a good flux reference for the intrinsic strength of \lya, because it arises in the low transition region of the upper stellar atmosphere at temperatures $\sim$50,000 K, overlapping the upper formation range of \lya\ itself \citep{linsky17-1}. After the scaling, the highest velocity wings of the \lya\ line profile ($> \pm500$~km~s$^{-1}$ from the line center) agree fairly well with the EK~Dra and DS~Tuc spectra, as do the \ion{N}{5} emission lines. However, the inner \lya\ wings of \targ\ show a clear deficit relative to both comparison stars, despite the similarity in the ages and effective temperatures. This deficit occurs well outside the spectral region expected to be affected by IS absorption ($\sim -20$ and +13 km s$^{-1}$; \S~\ref{sec:ism}).

The weakening of the \lya\ inner wings is unlikely to be an instrumental effect, because the STIS and COS profiles of the two comparison stars agree, and were obtained at roughly similar times (within years) as the \targ\ pointings. The wing depression is unlikely to be caused by enhanced ISM H I absorption toward the $3\times$ more distant HIP 67522, because the ISM-affected zone already is much narrower than the geocoronal exclusion region, and adding more column density would broaden the saturated IS absorption wavelengths only by roughly the square-root of the column density. More likely, the apparent extra \ion{H}{1} absorption is caused by a source of neutral hydrogen within the \targ\ system itself. Absorption local to the system, perhaps from material that has escaped from the exoplanet, could be an explanation \citep[e.g.,][]{haswell12-1,fossati15-1}. The exceedingly broad width of the absorption points to material that traces a hot component (if thermally broadened) or at high velocity. 

\begin{figure}
    \centering
    \includegraphics[width=0.7\textwidth]{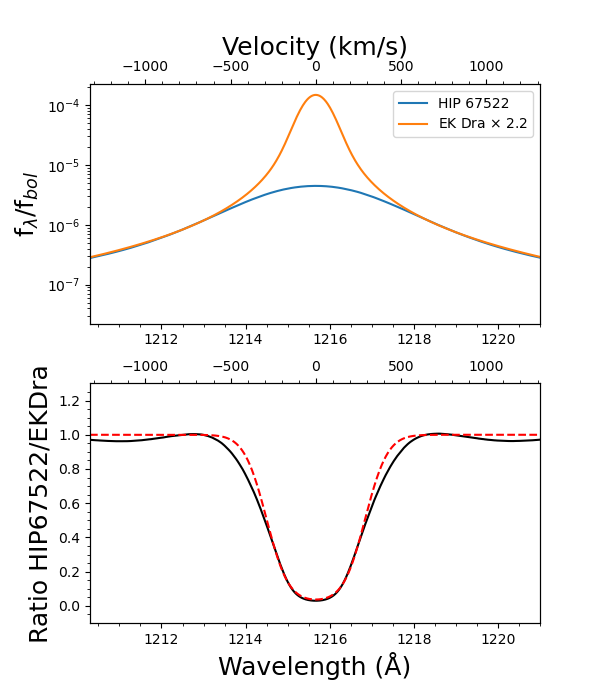}
    \caption{A comparison between HIP 67522's reconstructed \lya\ profile and EK Dra's (top) provides an estimate of the extra HI absorption present in the system (bottom). Top: HIP 67522's intrinsic \lya\ profile as estimated via the ``constrained" reconstruction is shown in blue and EK Dra's intrinsic \lya\ profile is shown in orange. Both profiles have been normalized by $f_{bol}$, and EK Dra's profile has been shifted in wavelength to match HIP 67522's radial velocity and multiplied by 2.2$\times$ to match HIP 67522's flux density beyond $>$500 km/s. Bottom: the ratio of the shifted and scaled profiles reveals a deep and broad absorption profile, representing an estimate of the extra absorption that may be present in the system. The ratio is shown in black, while the dotted red line is the fit to the residual absorption.}
    \label{fig:absorber}
\end{figure}

To roughly estimate the properties of this excess absorption, we compare the \targ\ intrinsic line profile from the ``constrained fit" \lya\ reconstruction with EK Dra's intrinsic line profile, which we reconstructed following the methods described previously (Figure~\ref{fig:absorber}). We shifted and scaled the EK Dra intrinsic profile to match HIP 67522’s intrinsic profile at $>$500 km/s from line center. Dividing the \targ\ profile by the EK~Dra profile, we obtain a broad, symmetric absorption profile centered in the stellar rest frame of \targ. The absorption profile is roughly fit by $\log N(HI) \sim 15$ and $b \sim  200$~km~s$^{-1}$.  This excess cannot be fit by the \lya\ reconstruction method because the high velocities and consequent broadening parameter fall outside the range attributable to interstellar absorption. Such a large Doppler broadening parameter would indicate  2--3~MK, coronal gas, possibly associated with the star, where most of the hydrogen is ionized and a small fraction is in neutral form.  However, this appears to be an unlikely scenario because such high temperature plasma containing highly ionized hydrogen 
would not produce significant Lyma alpha absorption.
Higher spectral resolution and/or transit observations of \targ, particularly imaging spectroscopy of \lya\ (to enable accurate geocoronal subtraction) or a tracer of similar excitation energies such as \ion{Mg}{2}, would allow a direct probe of the source of the excess \lya\ absorption in this system. Upcoming UV transit observations of HIP67522b (GO \#18121, PI Duvvuri) will look for evidence of mass loss from the planet in the system.

\subsection{The Sun in Time Revisited}

\begin{figure}
    \centering
    \includegraphics[width=0.9\textwidth]{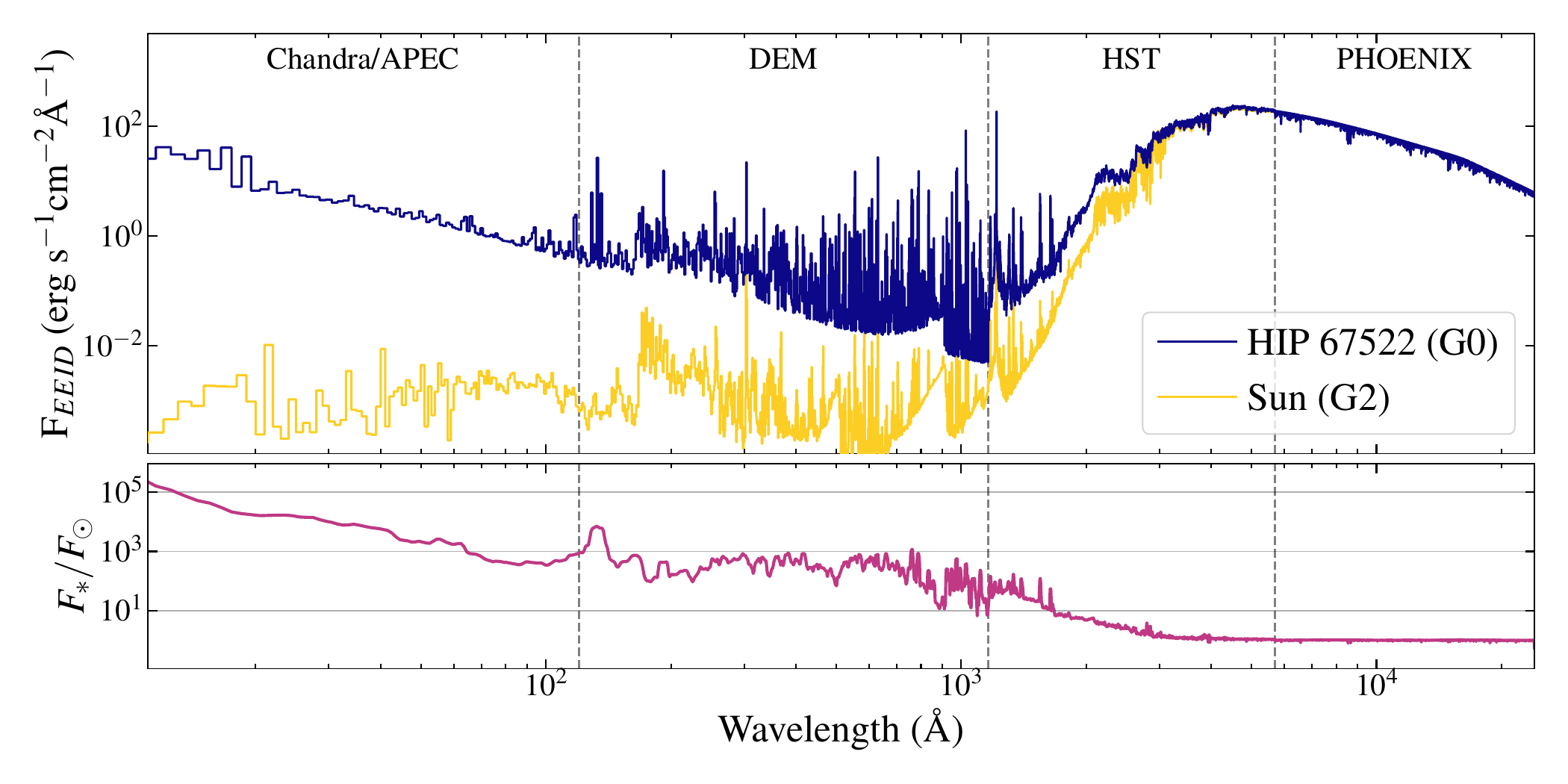}
\caption{The spectral energy distribution of \targ\ compared to the present-day Sun. Shown in blue in the upper panel is the SED of the 17~Myr old star, \targ. Overlaid in yellow is the SED of the Sun  from \citet{woods09-1}}. For \targ, the fluxes are scaled to an Earth-equivalent instellation distance. In the lower panel, the ratio of the two spectra is shown, displaying the dramatic increase in XUV emission in the younger star.  
    \label{fig:sed}
\end{figure}

In Figure~\ref{fig:sed}, we compare the broadband SED of \targ\ (first presented in \citet{thao24-1}) to the solar SED \citep{woods09-1}. \targ\ has been scaled to  an Earth-equivalent instellation distance, or $= 1 AU \times (L_{\star}/L_{\odot})^{0.5}$, which in this case is 1.32~AU. Overlying the comparable photospheric emission is a dramatic decrease in the upper atmospheric emission with age, such that \targ\ displays elevated X-ray to UV flux densities, exceeding solar values by 10$^{2}$--10$^{5}$ from the EUV to the shortest X-ray data observed ($\simeq5$~\AA). The comparison highlights the importance of the evolution of XUV emission from solar-type stars over their lifetimes when assessing exoplanet atmospheric properties and the potential for long-term atmospheric retention.

\begin{figure}
    \centering
    \includegraphics[width=0.9\textwidth]{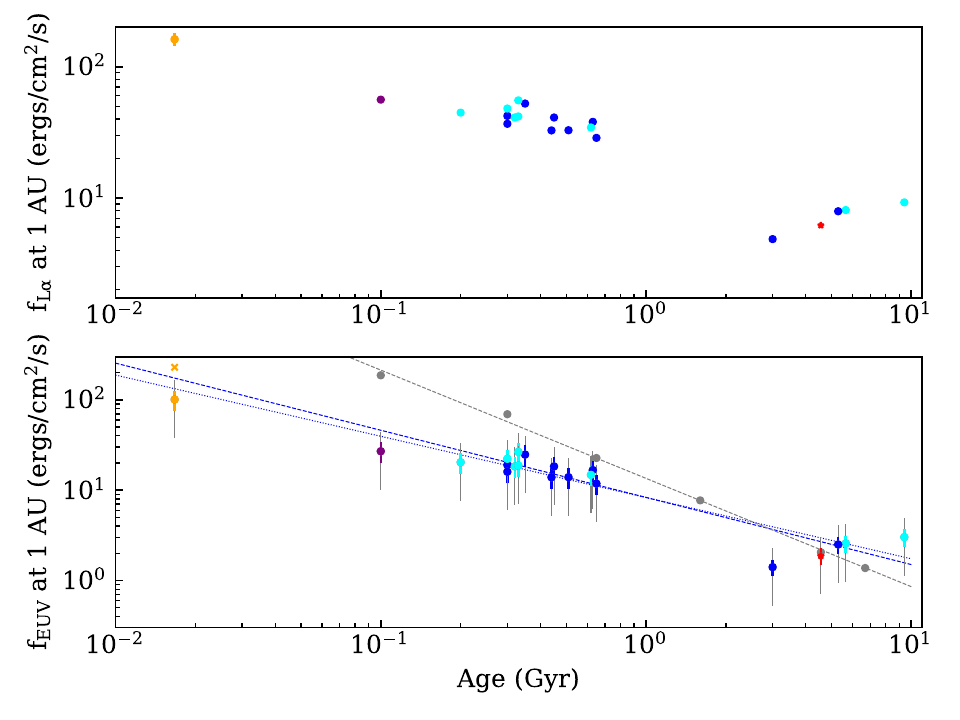}
    \caption{Shown in the top panel are the integrated \lya\ line fluxes for the stars given in Table~\ref{tab:euv}. The fluxes are normalized to a distance of 1 AU and the radius of a 1~M$_{\odot}$ star at that age.  \targ\ is plotted in orange.  The Sun is plotted as a red star.  EK~Dra is plotted in purple.} Stars shown in blue have spectral types from G0V to G5V, while stars shown in cyan have spectral types from G6V--G8V. Shown in the bottom panel is the  10--40~nm EUV flux density estimated using the scaling relations from \lya\ given in Table~5 of  \citet{linsky13-1}. The colored error bars on the EUV fluxes show the uncertainty from the scaling relation equations, while the light gray bars show the propagated deviation about the fits from Linsky et al. Plotted as an orange cross is the integrated EUV flux from the DEM model for \targ. Also plotted in gray are the  10--36~nm EUV flux densities and best fit  from \citet[][their Table 4]{ribas05-1}. The two blue lines show linear fits to the EUV fluxes estimated in this paper, with the dotted line showing the best fit when the EUV point for \targ\ is taken from the \lya\ scaling and the dashed line when the DEM flux is used. 
    \label{fig:euv}
\end{figure}

Since R05's influential examination of the high energy emission from the Sun in time, the number of high quality UV observations of solar type stars has grown. However, direct observations of EUV emission have not increased as there is currently no mission that can observe that waveband, so we must continue to estimate the EUV contribution from scaling relations and/or modeling. In Table~\ref{tab:euv} and Figure~\ref{fig:euv}, we present the current view of FUV (as represented by \lya, the dominant emission source) and EUV (100--400~\AA) emission from solar type stars as a function of their age. To allow direct comparison to R05, we plot fluxes at 1~AU and normalized the values to the expected radius of the Sun at that star's age.\footnote{For the latter, we used the values presented in R05 for the older stars and assumed a radius of 0.87~R$_{\sun}$, the Zero Age Main Sequence radius of the Sun from \citet{bressan2012} for \targ.} The \lya\ fluxes (aside from that of the Sun, which was directly measured) were all derived from \lya\ reconstruction based on HST UV spectra. The EUV emission is estimated using the scaling relation from \lya\ given by  \citet[][their Table~5]{linskyetal14-1}. The error bars for the \lya\ flux in \targ\ are taken from the scaling relation in \citet{woodetal05-1}, while the EUV error bars are calculated from the uncertainties in the Linsky et al.\ \lya\ to EUV scaling relations. Linsky et al.\ (their Section 4.4) note that the 20--37\% deviations they see about their fits are consistent with overall uncertainties in their method arising from the \lya\ reconstruction, measured EUV flux errors, and stellar variability. Our EUV calculations use three bands from their paper; propagation of the dispersions about their fits in each band give an overall uncertainty of 63\%, which is shown by the gray error on each point. The EUV emission from \targ\ calculated from our DEM is also shown as the upper orange point in the lower panel. The R05 EUV (10--36~nm) fluxes are presented in gray in that panel as is their powerlaw fit to their data. Finally, we also show later G-type stars (G6--G8V) in cyan to examine how high energy emission varies as a function of stellar effective temperature.

\begin{table}
    \centering
     \caption{\lya and EUV Fluxes}
    \label{tab:euv}
    \begin{tabular}{lccccl}
\hline \hline
Star & Spectral Type & Age & $F_{\mathrm{Ly\alpha}}$  & $F_{EUV}$  &  Ref. \\
 & & (Gyr) &  (erg s$^{-1}$ cm$^{-2}$) & (erg s$^{-1}$ cm$^{-2}$) \\
\hline 
HIP67522 & G0V & 0.017 & 162$\pm$18 & 101.0$\pm26$ & This paper\tablenotemark{a} \\
& & & & 230.2 & This paper (DEM value) \\
HR6748 & G0V & 0.44 & 32.7 & 13.9$\pm4$ & L20 \\
V993 Tau & G0V & 0.63 & 38.0 & 16.7$\pm4$ & L20 \\
EK Dra & G1.5V & 0.1 & 56.09 & 27.0$\pm7$ & S24  \\
Pi UMa & G2V & 0.3 & 42.2 & 19.0$\pm5$ & R05 \\
HR2882 & G2V & 0.35 & 52.4 & 24.8$\pm7$ & L20 \\
HR4345 & G2V & 0.45 & 41.0 & 18.3$\pm5$ & L20 \\
Chi Ori & G2V & 0.3 & 36.7 & 16.0$\pm4$ & L20 \\
Pi Men & G2V & 3.0 & 4.86 & 1.40$\pm0.3$ & L20 \\
Sun & G2V & 4.56 & 6.19 & 1.87$\pm0.4$ & R05 \\
SAO136111 & G5V & 0.51 & 32.8 & 13.9$\pm4$ & L20 \\
Kappa Cet & G5V & 0.65 & 28.7 & 11.9$\pm3$ & L20 \\
alpha Cen A & G5V & 5.3 & 7.9 & 2.50$\pm0.5$ & L20 \\
SAO158720 & G6V & 0.62 & 34.4 & 14.8$\pm4$ & L20 \\
HR2225 & G6V & 0.32 & 41.0 & 18.3$\pm5$ & L20\\
Chi Boo A & G7V & 0.2 & 44.7 & 20.4$\pm5$ & L20 \\
61 Vir & G7V & 9.41 & 9.26 & 3.01$\pm0.7$ & L20 \\
SAO254993 & G7V & 0.33 & 55.4 & 22.2$\pm7$ & L20 \\
HR 8 & G7V & 0.3 & 48.0 & 22.2$\pm26$ & L20 \\
SAO28753 & G7V & 0.33 & 41.8 & 18.8$\pm5$ & L20 \\
Tau Ceti & G8V & 5.66 & 8.09 & 2.57$\pm0.5$ & L20 \\
\hline
   \end{tabular}
\tablenotetext{a}{Taken from the ``Stellar Rotation Scaling" in Table~\ref{tab:lines}.}   
\tablecomments{$F_{\mathrm{Ly\alpha}}$ is the \lya\ flux at 1~AU in each system. $F_{EUV}$ is the 100--400~\AA\ flux at 1~AU and normalized to the expected radius of the Sun at the age of the system to be consistent with the values presented in Ribas et al.\ (2005). } References: R05--Ribas et al.\ 2005; L20--Linsky et al.\ 2020, S24-- Shoda et al.\ 2024\nocite{shoda24-1}  

\end{table}

The evolution of \lya\ with age exhibits a steady decrease, consistent with the results from \citet{linskyetal20-1}, who found that $L_{
Ly_{\alpha}}/L_{bolometric}$ decreased smoothly with age---here, we see this trend extended to the earliest lifetime of the star.  The EUV emission is directly scaled to the \lya\ value and thus shows the same variation.  We fit our EUV fluxes using a linear fit of the form $\log(F_{EUV}) = m \times log(Age) +b$, where the age is in Gyr and the fluxes in ergs~cm$^-2$~s$^{-1}$ at 1~AU. In Figure~\ref{fig:euv} we show fits both for the \lya\ scaling relation-derived flux for \targ\ and the DEM value. Our results give $m = -0.68\pm0.05$, $b=0.918\pm0.036$ and $m = -0.75\pm0.05$, $b=0.919\pm0.036$, respectively; the differences between the two fits are statistically insignificant. 

The index of the resulting power law in Figure~\ref{fig:euv} is -0.68, significantly more shallow than the -1.12 power law index found by R05, as our EUV points increasingl deviate from those of R05 to younger ages. Our scaling relation calculation for the EUV flux of EK~Dra is a factor of 7 below the observationally-derived value from the R05, for example. A recent analysis (Youngblood et al.\ 2025, in press) finds an interstellar column density to EK~Dra that is lower than that used by R05 in their data analysis ($\log(N(HI) = 18.04$ vs $\log(N(HI) = 18.18$ used by R05), but this would only decrease the R05 flux by $\sim$20\%. One possible explanation for the discrepency may be that the EUV emission correlation with \lya\ underestimates the contribution from the transition region and corona.  In a companion paper, \citet{france25-1}, we examine in more detail the evolution of the EUV flux of Sun-like stars over time using various scaling relations. 

For \targ, \citet{maggio24-1} used an emission measure distribution based on X-ray and FUV line to calculate a XUV (5--920~\AA) flux of $3.9\times10^{5}$~ergs~cm$^{-2}$~s$^{-1}$ at the planet HIP67522b (0.076~AU). Our calculations based on our DEM  gives a lower value,  $1.0\times10^{5}$~ergs~cm$^{-2}$~s$^{-1}$, but within a factor of four.

\subsection{The Fate of Close-in Planets in \targ}

As shown in Figure~\ref{fig:sed}, our SED predicts that \targ\ emits 100-1000 times the EUV flux density of the Sun, rising to 10$^{5}$ in X-rays. This early intense irradiation, coupled with the larger radius/lower density of young planets due to residual heat from their formation, makes the early (10--300~Myr) lifetime of the system the key epoch for atmospheric escape from close-in, gaseous planets. The duration of this phase is set by the initial rotation rate of the star \citep{tu15-1,johnstone21-1}. The total material loss depends on the mass of the planet and the evolutionary timescale of the stellar evolution of upper atmosphere. The mass loss modeling performed by \citet{maggio24-1} for HIP67522b  predicts a mass loss rate of $\sim10^{-2}$~M$_{\earth}$~Myr$^{-1}$ and an e-folding evaporation rate of 300--600~Myr for a 57~M$_{\earth}$ planet. Their analysis preceded the transit spectroscopy results of \citet{thao24-1}, which showed that HIP~67522b is  $\simeq4\times$ less massive/less dense than the Maggio et al.\ low mass case, which will drive the evaporation timescale even shorter than they determined. 

From our SED, we can calculate the expected relative contributions of X-ray (1--100~\AA) and EUV (100--900~\AA): at 1~AU , they contribute 329 and 672~ergs~s$^{-1}$~cm$^{-2}$, respectively, while at the distance of HIP67522~b (0.076~AU), the values are 57,000 and 116,000~ergs~s$^{-1}$~cm$^{-2}$.  \citet{owen12-1} present a planetary escape model that is dominated by X-ray driven escape for close-in, Neptune-mass planets. Interpolating from their Figure 11, we find a predicted transition from  X-ray-dominated to EUV-dominated escape for a planet with the location and density of HIP~67522b occurs at a X-ray luminosity of $\sim10^{29}$~ergs~s$^{-1}$. We measured $L_{X} = 2.2\times10^{30}$~ergs~s$^{-1}$, which would suggest that atmospheric escape in HIP67522b is dominated by X-rays. However, Owen et al.\ assumed that the X-ray and EUV emission contributions were equal, whereas we find a factor of two increase in flux as one transitions to the EUV. The exact breakdown of X-ray and EUV components to planetary escape in HIP67522b will require more specific modeling to resolve.

In their analysis of the evolution of the Sun, \citet{johnstone21-2} found that the Sun would need to have been born a slow to moderate rotator ($>4$~d initial rotation period) to prevent rapid escape of Earth's atmosphere, with its high activity level decaying early, within the first 20--50~Myr. \targ\ is a rapid rotator (P = 1.4~d) and its two giant planets lie interior to Earth's orbit, at 0.078 and 0.122~AU. By observing the earliest stages of a more active counterpart to the Sun, we are seeing an alternate evolutionary pathway for a solar type star in which strong XUV emission is poised to strip the envelopes of its close-in gaseous planets.

\section{Conclusion} \label{sec:conclusion}

In this paper, we summarized the XUV characteristics of the youngest solar-type planet with transiting exoplanet companions at the epoch when XUV activity and interior planet heating combine to maximize the potential for atmospheric escape. \targ\ has two transiting planets and one, HIP67522b, has been shown to be a highly inflated, low density precursor to the sub-Neptunes seen in older systems. We find that the XUV emission flux density exceeds present-day solar levels by $10^{2} - 10^{5}$ and is characterized by continuous low-level variability and regular flaring. Our usual method of reconstructing the intrinsic \lya\ emission from the star returned unphysical values. The explanation appears to be an unknown source of extra absorption of \ion{H}{1} that extends to $> \pm500$~km~s$^{-1}$  intrinsic to the system. Local absorption due to escaped gas from the exoplanet(s) may be a cause, although higher spectral resolution and/or transit observations of \lya\ and \ion{Mg}{2} would be needed to confirm. We reexamine the picture of solar-type stars in time, showing that chromospheric \lya\ continues to smoothly vary with stellar age, while the EUV component shows signs of saturating at $<100$~Myr. The initial rapid rotation period of the star, coupled with the low density of the close-in planets present a picture of an alternate Sun that will evolve into rocky planets without atmospheres under the extended XUV saturation phase.


\acknowledgments
We thank the anonymous referee for careful, thoughtful attention to the paper that significantly improved the final manuscript.Based on observations made with the NASA/ESA Hubble Space Telescope, obtained from the Data Archive at the Space Telescope Science Institute, which is operated by the Association of Universities for Research in Astronomy, Inc., under NASA contract NAS 5-26555. These observations are associated with program \# 16701. Support for program \#16701 was provided by NASA through a grant from the Space Telescope Science Institute, which is operated by the Association of Universities for Research in Astronomy, Inc., under NASA contract NAS 5-26555. All of the \textit{HST} data presented in this paper were obtained from the Mikulski Archive for Space Telescopes (MAST). This research employs a list of Chandra datasets, obtained by the Chandra X-ray Observatory, contained in~\dataset[DOI: 10.25574/cdc.471]{https://doi.org/10.25574/cdc.471}."

\appendix
Shown below are the corner plots for the \lya\ reconstruction analysis presented in Section~\ref{sec:lya}.

\begin{figure}
    \centering
     \includegraphics[width=0.9\textwidth]{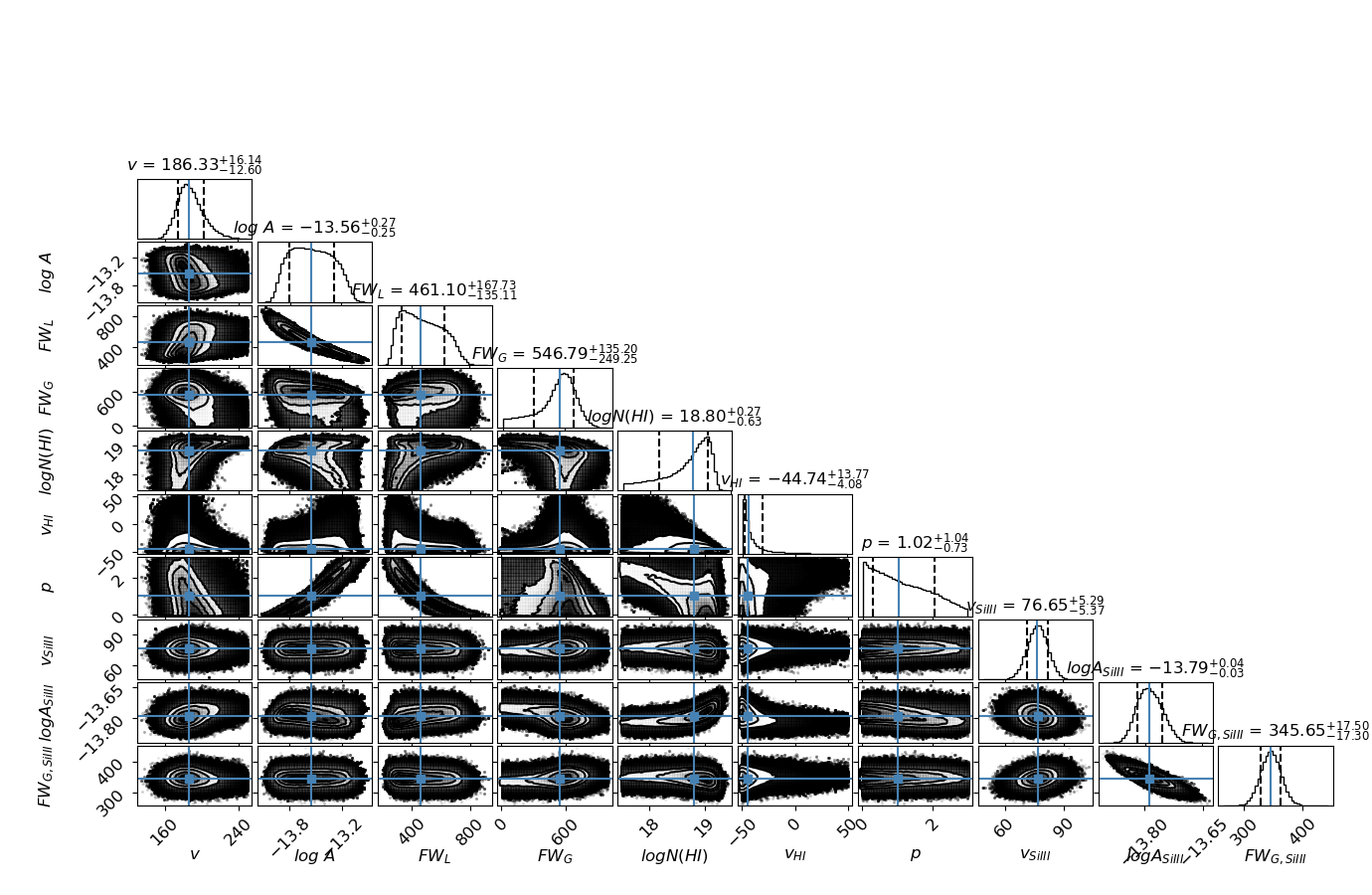}
    \caption{Corner plot for the left panel fit shown in Figure~\ref{fig:lya1}. One- and two-dimensional projections of the sampled posterior probability distributions, referred to as marginalized and joint distributions, respectively, of the nine parameters for the  fit. Contours in the joint distributions are shown at 0.5, 1, 1.5, and 2$\sigma$, and the histograms' dashed black vertical lines show the 16th, 50th, and 84th percentiles of the samples in each marginalized distribution. Text above each histogram shows the median $\pm$ the 68\% confidence interval.}
    \label{fig:lya1corner}
\end{figure}

\begin{figure}
    \centering
    \includegraphics[width=0.9\textwidth]{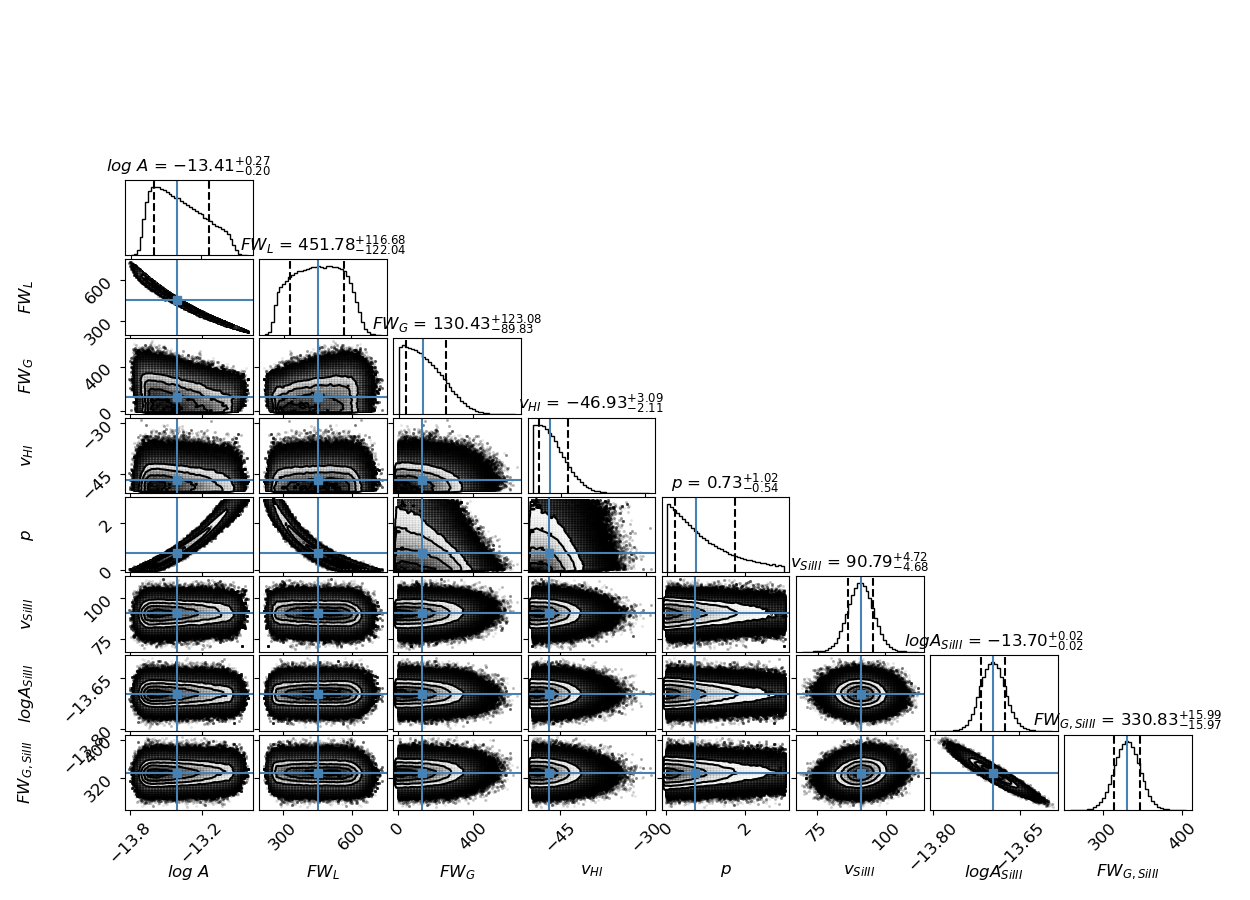}
    \caption{Corner plot for right panel fit shown in Figure~\ref{fig:lya1}. }
    \label{fig:lya2corner}
\end{figure}



\vspace{5mm}
\facilities{\textit{HST} (COS,STIS),\textit{Chandra},   \textit{TESS},\textit{eROSITA} }

\defcitealias{astropy18-1}{Astropy Collaboration, 2018}
\software{astropy \citepalias{astropy18-1}, xspec \citep{arnaud96-1}, stistools\footnote{\url{https://stistools.readthedocs.io/en/latest/}}, scipy \citep{virtanenetal20-1}, numpy \citep{harrisetal20-1}, matplotlib \citep{hunter07-1}}

\bibliographystyle{aasjournal}

\bibliography{aabib,newcites}

\begin{thebibliography}{}
\expandafter\ifx\csname natexlab\endcsname\relax\def\natexlab#1{#1}\fi
\providecommand{\url}[1]{\href{#1}{#1}}
\providecommand{\dodoi}[1]{doi:~\href{http://doi.org/#1}{\nolinkurl{#1}}}
\providecommand{\doeprint}[1]{\href{http://ascl.net/#1}{\nolinkurl{http://ascl.net/#1}}}
\providecommand{\doarXiv}[1]{\href{https://arxiv.org/abs/#1}{\nolinkurl{https://arxiv.org/abs/#1}}}

\bibitem[{{Arnaud}(1996)}]{arnaud96-1}
{Arnaud}, K.~A. 1996, in Astronomical Society of the Pacific Conference Series,
  Vol. 101, Astronomical Data Analysis Software and Systems V, ed. G.~H.
  {Jacoby} \& J.~{Barnes}, 17

\bibitem[{{Astropy Collaboration} {et~al.}(2018){Astropy Collaboration},
  {Price-Whelan}, {Sip{\H{o}}cz}, {G{\"u}nther}, {Lim}, {Crawford}, {Conseil},
  {Shupe}, {Craig}, {Dencheva}, {Ginsburg}, {Vand erPlas}, {Bradley},
  {P{\'e}rez-Su{\'a}rez}, {de Val-Borro}, {Aldcroft}, {Cruz}, {Robitaille},
  {Tollerud}, {Ardelean}, {Babej}, {Bach}, {Bachetti}, {Bakanov}, {Bamford},
  {Barentsen}, {Barmby}, {Baumbach}, {Berry}, {Biscani}, {Boquien}, {Bostroem},
  {Bouma}, {Brammer}, {Bray}, {Breytenbach}, {Buddelmeijer}, {Burke},
  {Calderone}, {Cano Rodr{\'\i}guez}, {Cara}, {Cardoso}, {Cheedella}, {Copin},
  {Corrales}, {Crichton}, {D'Avella}, {Deil}, {Depagne}, {Dietrich}, {Donath},
  {Droettboom}, {Earl}, {Erben}, {Fabbro}, {Ferreira}, {Finethy}, {Fox},
  {Garrison}, {Gibbons}, {Goldstein}, {Gommers}, {Greco}, {Greenfield},
  {Groener}, {Grollier}, {Hagen}, {Hirst}, {Homeier}, {Horton}, {Hosseinzadeh},
  {Hu}, {Hunkeler}, {Ivezi{\'c}}, {Jain}, {Jenness}, {Kanarek}, {Kendrew},
  {Kern}, {Kerzendorf}, {Khvalko}, {King}, {Kirkby}, {Kulkarni}, {Kumar},
  {Lee}, {Lenz}, {Littlefair}, {Ma}, {Macleod}, {Mastropietro}, {McCully},
  {Montagnac}, {Morris}, {Mueller}, {Mumford}, {Muna}, {Murphy}, {Nelson},
  {Nguyen}, {Ninan}, {N{\"o}the}, {Ogaz}, {Oh}, {Parejko}, {Parley}, {Pascual},
  {Patil}, {Patil}, {Plunkett}, {Prochaska}, {Rastogi}, {Reddy Janga},
  {Sabater}, {Sakurikar}, {Seifert}, {Sherbert}, {Sherwood-Taylor}, {Shih},
  {Sick}, {Silbiger}, {Singanamalla}, {Singer}, {Sladen}, {Sooley},
  {Sornarajah}, {Streicher}, {Teuben}, {Thomas}, {Tremblay}, {Turner},
  {Terr{\'o}n}, {van Kerkwijk}, {de la Vega}, {Watkins}, {Weaver}, {Whitmore},
  {Woillez}, {Zabalza}, \& {Astropy Contributors}}]{astropy18-1}
{Astropy Collaboration}, {Price-Whelan}, A.~M., {Sip{\H{o}}cz}, B.~M., {et~al.}
  2018, \aj, 156, 123, \dodoi{10.3847/1538-3881/aabc4f}

\bibitem[{{Ayres}(1997)}]{ayres97-1}
{Ayres}, T.~R. 1997, \jgr, 102, 1641, \dodoi{10.1029/96JE03306}

\bibitem[{{Ayres}(2022)}]{ayres22-1}
---. 2022, \aj, 163, 78, \dodoi{10.3847/1538-3881/ac3762}

\bibitem[{Barber {et~al.}(2024)Barber, Thao, Mann, Vanderburg, Mori,
  Livingston, Fukui, Narita, Kraus, Tofflemire, Newton, Winn, Jenkins, Seager,
  Collins, \& Twicken}]{barber24-1}
Barber, M.~G., Thao, P.~C., Mann, A.~W., {et~al.} 2024, The Astrophysical
  Journal Letters, 973, L30, \dodoi{10.3847/2041-8213/ad77d9}

\bibitem[{{Bourrier} {et~al.}(2017){Bourrier}, {Ehrenreich}, {Wheatley},
  {Bolmont}, {Gillon}, {de Wit}, {Burgasser}, {Jehin}, {Queloz}, \&
  {Triaud}}]{bourrieretal17-1}
{Bourrier}, V., {Ehrenreich}, D., {Wheatley}, P.~J., {et~al.} 2017, \aap, 599,
  L3, \dodoi{10.1051/0004-6361/201630238}

\bibitem[{{Bourrier} {et~al.}(2018){Bourrier}, {Lecavelier des Etangs},
  {Ehrenreich}, {Sanz-Forcada}, {Allart}, {Ballester}, {Buchhave}, {Cohen},
  {Deming}, {Evans}, {Garc{\'\i}a Mu{\~n}oz}, {Henry}, {Kataria}, {Lavvas},
  {Lewis}, {L{\'o}pez-Morales}, {Marley}, {Sing}, \&
  {Wakeford}}]{bourrieretal18-1}
{Bourrier}, V., {Lecavelier des Etangs}, A., {Ehrenreich}, D., {et~al.} 2018,
  \aap, 620, A147, \dodoi{10.1051/0004-6361/201833675}

\bibitem[{{Bowyer} \& {Malina}(1991)}]{bowyer91-1}
{Bowyer}, S., \& {Malina}, R.~F. 1991, Advances in Space Research, 11, 205,
  \dodoi{10.1016/0273-1177(91)90077-W}

\bibitem[{{Bressan} {et~al.}(2012){Bressan}, {Marigo}, {Girardi}, {Salasnich},
  {Dal Cero}, {Rubele}, \& {Nanni}}]{bressan2012}
{Bressan}, A., {Marigo}, P., {Girardi}, L., {et~al.} 2012, \mnras, 427, 127,
  \dodoi{10.1111/j.1365-2966.2012.21948.x}

\bibitem[{{Brown} {et~al.}(2023){Brown}, {Schneider}, {France}, {Froning},
  {Youngblood}, {J. Wilson}, {Loyd}, {Pineda}, {Duvvuri}, {Kowalski}, \&
  {Berta-Thompson}}]{brown23-1}
{Brown}, A., {Schneider}, P.~C., {France}, K., {et~al.} 2023, \aj, 165, 195,
  \dodoi{10.3847/1538-3881/acc38a}

\bibitem[{{Foreman-Mackey} {et~al.}(2013){Foreman-Mackey}, {Hogg}, {Lang}, \&
  {Goodman}}]{foreman-mackeyetal13-1}
{Foreman-Mackey}, D., {Hogg}, D.~W., {Lang}, D., \& {Goodman}, J. 2013, \pasp,
  125, 306, \dodoi{10.1086/670067}

\bibitem[{{Fossati} {et~al.}(2015){Fossati}, {France}, {Koskinen}, {Juvan},
  {Haswell}, \& {Lendl}}]{fossati15-1}
{Fossati}, L., {France}, K., {Koskinen}, T., {et~al.} 2015, \apj, 815, 118,
  \dodoi{10.1088/0004-637X/815/2/118}

\bibitem[{{France} {et~al.}(2018){France}, {Arulanantham}, {Fossati}, {Lanza},
  {Loyd}, {Redfield}, \& {Schneider}}]{franceetal18-1}
{France}, K., {Arulanantham}, N., {Fossati}, L., {et~al.} 2018, \apjs, 239, 16,
  \dodoi{10.3847/1538-4365/aae1a3}

\bibitem[{{France} {et~al.}(2013){France}, {Froning}, {Linsky}, {Roberge},
  {Stocke}, {Tian}, {Bushinsky}, {D{\'e}sert}, {Mauas}, {Vieytes}, \&
  {Walkowicz}}]{franceetal13-1}
{France}, K., {Froning}, C.~S., {Linsky}, J.~L., {et~al.} 2013, \apj, 763, 149,
  \dodoi{10.1088/0004-637X/763/2/149}

\bibitem[{{France} {et~al.}(2025){France}, {Duvvuri}, {Froning}, {Brown},
  {Schneider}, {Pineda}, {Wilson}, {Youngblood}, {Airapetian}, {Namekata},
  {Notsu}, \& {Sextro}}]{france25-1}
{France}, K., {Duvvuri}, G., {Froning}, C.~S., {et~al.} 2025, \aj, 170, 159,
  \dodoi{10.3847/1538-3881/adefdf}

\bibitem[{{Fulton} {et~al.}(2017){Fulton}, {Petigura}, {Howard}, {Isaacson},
  {Marcy}, {Cargile}, {Hebb}, {Weiss}, {Johnson}, {Morton}, {Sinukoff},
  {Crossfield}, \& {Hirsch}}]{fulton17-1}
{Fulton}, B.~J., {Petigura}, E.~A., {Howard}, A.~W., {et~al.} 2017, \aj, 154,
  109, \dodoi{10.3847/1538-3881/aa80eb}

\bibitem[{{Gaia Collaboration} {et~al.}(2023){Gaia Collaboration}, {Vallenari},
  {Brown}, {Prusti}, {de Bruijne}, {Arenou}, {Babusiaux}, {Biermann},
  {Creevey}, {Ducourant}, {Evans}, {Eyer}, {Guerra}, {Hutton}, {Jordi},
  {Klioner}, {Lammers}, {Lindegren}, {Luri}, {Mignard}, {Panem}, {Pourbaix},
  {Randich}, {Sartoretti}, {Soubiran}, {Tanga}, {Walton}, {Bailer-Jones},
  {Bastian}, {Drimmel}, {Jansen}, {Katz}, {Lattanzi}, {van Leeuwen}, {Bakker},
  {Cacciari}, {Casta{\~n}eda}, {De Angeli}, {Fabricius}, {Fouesneau},
  {Fr{\'e}mat}, {Galluccio}, {Guerrier}, {Heiter}, {Masana}, {Messineo},
  {Mowlavi}, {Nicolas}, {Nienartowicz}, {Pailler}, {Panuzzo}, {Riclet}, {Roux},
  {Seabroke}, {Sordo}, {Th{\'e}venin}, {Gracia-Abril}, {Portell}, {Teyssier},
  {Altmann}, {Andrae}, {Audard}, {Bellas-Velidis}, {Benson}, {Berthier},
  {Blomme}, {Burgess}, {Busonero}, {Busso}, {C{\'a}novas}, {Carry}, {Cellino},
  {Cheek}, {Clementini}, {Damerdji}, {Davidson}, {de Teodoro}, {Nu{\~n}ez
  Campos}, {Delchambre}, {Dell'Oro}, {Esquej}, {Fern{\'a}ndez-Hern{\'a}ndez},
  {Fraile}, {Garabato}, {Garc{\'\i}a-Lario}, {Gosset}, {Haigron}, {Halbwachs},
  {Hambly}, {Harrison}, {Hern{\'a}ndez}, {Hestroffer}, {Hodgkin}, {Holl},
  {Jan{\ss}en}, {Jevardat de Fombelle}, {Jordan}, {Krone-Martins}, {Lanzafame},
  {L{\"o}ffler}, {Marchal}, {Marrese}, {Moitinho}, {Muinonen}, {Osborne},
  {Pancino}, {Pauwels}, {Recio-Blanco}, {Reyl{\'e}}, {Riello}, {Rimoldini},
  {Roegiers}, {Rybizki}, {Sarro}, {Siopis}, {Smith}, {Sozzetti}, {Utrilla},
  {van Leeuwen}, {Abbas}, {{\'A}brah{\'a}m}, {Abreu Aramburu}, {Aerts},
  {Aguado}, {Ajaj}, {Aldea-Montero}, {Altavilla}, {{\'A}lvarez}, {Alves},
  {Anders}, {Anderson}, {Anglada Varela}, {Antoja}, {Baines}, {Baker},
  {Balaguer-N{\'u}{\~n}ez}, {Balbinot}, {Balog}, {Barache}, {Barbato},
  {Barros}, {Barstow}, {Bartolom{\'e}}, {Bassilana}, {Bauchet}, {Becciani},
  {Bellazzini}, {Berihuete}, {Bernet}, {Bertone}, {Bianchi}, {Binnenfeld},
  {Blanco-Cuaresma}, {Blazere}, {Boch}, {Bombrun}, {Bossini}, {Bouquillon},
  {Bragaglia}, {Bramante}, {Breedt}, {Bressan}, {Brouillet}, {Brugaletta},
  {Bucciarelli}, {Burlacu}, {Butkevich}, {Buzzi}, {Caffau}, {Cancelliere},
  {Cantat-Gaudin}, {Carballo}, {Carlucci}, {Carnerero}, {Carrasco},
  {Casamiquela}, {Castellani}, {Castro-Ginard}, {Chaoul}, {Charlot}, {Chemin},
  {Chiaramida}, {Chiavassa}, {Chornay}, {Comoretto}, {Contursi}, {Cooper},
  {Cornez}, {Cowell}, {Crifo}, {Cropper}, {Crosta}, {Crowley}, {Dafonte},
  {Dapergolas}, {David}, {David}, {de Laverny}, {De Luise}, \& {De
  March}}]{vallenari23-1}
{Gaia Collaboration}, {Vallenari}, A., {Brown}, A.~G.~A., {et~al.} 2023, \aap,
  674, A1, \dodoi{10.1051/0004-6361/202243940}

\bibitem[{Gordon(2024)}]{Gordon2024}
Gordon, K.~D. 2024, Journal of Open Source Software, 9, 7023,
  \dodoi{10.21105/joss.07023}

\bibitem[{{Gordon} {et~al.}(2023){Gordon}, {Clayton}, {Decleir}, {Fitzpatrick},
  {Massa}, {Misselt}, \& {Tollerud}}]{gordon23-1}
{Gordon}, K.~D., {Clayton}, G.~C., {Decleir}, M., {et~al.} 2023, \apj, 950, 86,
  \dodoi{10.3847/1538-4357/accb59}

\bibitem[{{G{\"u}del} {et~al.}(1997{\natexlab{a}}){G{\"u}del}, {Guinan},
  {Mewe}, {Kaastra}, \& {Skinner}}]{gudel97-1}
{G{\"u}del}, M., {Guinan}, E.~F., {Mewe}, R., {Kaastra}, J.~S., \& {Skinner},
  S.~L. 1997{\natexlab{a}}, \apj, 479, 416, \dodoi{10.1086/303859}

\bibitem[{{G{\"u}del} {et~al.}(1997{\natexlab{b}}){G{\"u}del}, {Guinan}, \&
  {Skinner}}]{gudel97-2}
{G{\"u}del}, M., {Guinan}, E.~F., \& {Skinner}, S.~L. 1997{\natexlab{b}}, \apj,
  483, 947, \dodoi{10.1086/304264}

\bibitem[{Harris {et~al.}(2020)Harris, Millman, van~der Walt, Gommers,
  Virtanen, Cournapeau, Wieser, Taylor, Berg, Smith, Kern, Picus, Hoyer, van
  Kerkwijk, Brett, Haldane, del R{'{\i}}o, Wiebe, Peterson,
  G{'{e}}rard-Marchant, Sheppard, Reddy, Weckesser, Abbasi, Gohlke, \&
  Oliphant}]{harrisetal20-1}
Harris, C.~R., Millman, K.~J., van~der Walt, S.~J., {et~al.} 2020, Nature, 585,
  357, \dodoi{10.1038/s41586-020-2649-2}

\bibitem[{{Haswell} {et~al.}(2012){Haswell}, {Fossati}, {Ayres}, {France},
  {Froning}, {Holmes}, {Kolb}, {Busuttil}, {Street}, {Hebb}, {Collier Cameron},
  {Enoch}, {Burwitz}, {Rodriguez}, {West}, {Pollacco}, {Wheatley}, \&
  {Carter}}]{haswell12-1}
{Haswell}, C.~A., {Fossati}, L., {Ayres}, T., {et~al.} 2012, \apj, 760, 79,
  \dodoi{10.1088/0004-637X/760/1/79}

\bibitem[{Hunter(2007)}]{hunter07-1}
Hunter, J.~D. 2007, Computing in Science \& Engineering, 9, 90,
  \dodoi{10.1109/MCSE.2007.55}

\bibitem[{{Jackson} {et~al.}(2012){Jackson}, {Davis}, \&
  {Wheatley}}]{jackson12-1}
{Jackson}, A.~P., {Davis}, T.~A., \& {Wheatley}, P.~J. 2012, \mnras, 422, 2024,
  \dodoi{10.1111/j.1365-2966.2012.20657.x}

\bibitem[{{Johnstone} {et~al.}(2021{\natexlab{a}}){Johnstone}, {Bartel}, \&
  {G{\"u}del}}]{johnstone21-1}
{Johnstone}, C.~P., {Bartel}, M., \& {G{\"u}del}, M. 2021{\natexlab{a}}, \aap,
  649, A96, \dodoi{10.1051/0004-6361/202038407}

\bibitem[{{Johnstone} {et~al.}(2021{\natexlab{b}}){Johnstone}, {Lammer},
  {Kislyakova}, {Scherf}, \& {G{\"u}del}}]{johnstone21-2}
{Johnstone}, C.~P., {Lammer}, H., {Kislyakova}, K.~G., {Scherf}, M., \&
  {G{\"u}del}, M. 2021{\natexlab{b}}, Earth and Planetary Science Letters, 576,
  117197, \dodoi{10.1016/j.epsl.2021.117197}

\bibitem[{{King} \& {Wheatley}(2021)}]{king21-1}
{King}, G.~W., \& {Wheatley}, P.~J. 2021, \mnras, 501, L28,
  \dodoi{10.1093/mnrasl/slaa186}

\bibitem[{{Kubyshkina} {et~al.}(2019){Kubyshkina}, {Cubillos}, {Fossati},
  {Erkaev}, {Johnstone}, {Kislyakova}, {Lammer}, {Lendl}, {Odert}, \&
  {G{\"u}del}}]{kubyshkina19-1}
{Kubyshkina}, D., {Cubillos}, P.~E., {Fossati}, L., {et~al.} 2019, \apj, 879,
  26, \dodoi{10.3847/1538-4357/ab1e42}

\bibitem[{{Lammer} {et~al.}(2003){Lammer}, {Selsis}, {Ribas}, {Guinan},
  {Bauer}, \& {Weiss}}]{lammer03-1}
{Lammer}, H., {Selsis}, F., {Ribas}, I., {et~al.} 2003, \apjl, 598, L121,
  \dodoi{10.1086/380815}

\bibitem[{{Linsky}(2017)}]{linsky17-1}
{Linsky}, J.~L. 2017, \araa, 55, 159,
  \dodoi{10.1146/annurev-astro-091916-055327}

\bibitem[{{Linsky} {et~al.}(2014){Linsky}, {Fontenla}, \&
  {France}}]{linskyetal14-1}
{Linsky}, J.~L., {Fontenla}, J., \& {France}, K. 2014, \apj, 780, 61,
  \dodoi{10.1088/0004-637X/780/1/61}

\bibitem[{{Linsky} {et~al.}(2013){Linsky}, {France}, \& {Ayres}}]{linsky13-1}
{Linsky}, J.~L., {France}, K., \& {Ayres}, T. 2013, \apj, 766, 69,
  \dodoi{10.1088/0004-637X/766/2/69}

\bibitem[{{Linsky} {et~al.}(2020){Linsky}, {Wood}, {Youngblood}, {Brown},
  {Froning}, {France}, {Buccino}, {Cranmer}, {Mauas}, {Miguel}, {Pineda},
  {Rugheimer}, {Vieytes}, {Wheatley}, \& {Wilson}}]{linskyetal20-1}
{Linsky}, J.~L., {Wood}, B.~E., {Youngblood}, A., {et~al.} 2020, \apj, 902, 3,
  \dodoi{10.3847/1538-4357/abb36f}

\bibitem[{{Lodders}(2003)}]{lodders03-1}
{Lodders}, K. 2003, \apj, 591, 1220, \dodoi{10.1086/375492}

\bibitem[{{Luhman}(2022)}]{luhman22-1}
{Luhman}, K.~L. 2022, \aj, 164, 151, \dodoi{10.3847/1538-3881/ac85e2}

\bibitem[{{Maggio} {et~al.}(2024){Maggio}, {Pillitteri}, {Argiroffi}, {Locci},
  {Benatti}, \& {Micela}}]{maggio24-1}
{Maggio}, A., {Pillitteri}, I., {Argiroffi}, C., {et~al.} 2024, \aap, 690,
  A383, \dodoi{10.1051/0004-6361/202451582}

\bibitem[{{Nisak} {et~al.}(2025){Nisak}, {Redfield}, {Linsky}, {Wood}, \&
  {Youngblood}}]{nisak25-1}
{Nisak}, A.~H., {Redfield}, S., {Linsky}, J.~L., {Wood}, B.~E., \&
  {Youngblood}, A. 2025, \apj, 983, 5, \dodoi{10.3847/1538-4357/adb960}

\bibitem[{{Owen} \& {Jackson}(2012)}]{owen12-1}
{Owen}, J.~E., \& {Jackson}, A.~P. 2012, \mnras, 425, 2931,
  \dodoi{10.1111/j.1365-2966.2012.21481.x}

\bibitem[{{Owen} \& {Wu}(2013)}]{owen13-1}
{Owen}, J.~E., \& {Wu}, Y. 2013, \apj, 775, 105,
  \dodoi{10.1088/0004-637X/775/2/105}

\bibitem[{{Owen} \& {Wu}(2017)}]{owen17-1}
---. 2017, \apj, 847, 29, \dodoi{10.3847/1538-4357/aa890a}

\bibitem[{{Pizzolato} {et~al.}(2003){Pizzolato}, {Maggio}, {Micela},
  {Sciortino}, \& {Ventura}}]{pizzolato03-1}
{Pizzolato}, N., {Maggio}, A., {Micela}, G., {Sciortino}, S., \& {Ventura}, P.
  2003, \aap, 397, 147, \dodoi{10.1051/0004-6361:20021560}

\bibitem[{{Ribas} {et~al.}(2005){Ribas}, {Guinan}, {G{\"u}del}, \&
  {Audard}}]{ribas05-1}
{Ribas}, I., {Guinan}, E.~F., {G{\"u}del}, M., \& {Audard}, M. 2005, \apj, 622,
  680, \dodoi{10.1086/427977}

\bibitem[{{Rizzuto} {et~al.}(2020){Rizzuto}, {Newton}, {Mann}, {Tofflemire},
  {Vanderburg}, {Kraus}, {Wood}, {Quinn}, {Zhou}, {Thao}, {Law}, {Ziegler}, \&
  {Brice{\~n}o}}]{rizzutoetal20-1}
{Rizzuto}, A.~C., {Newton}, E.~R., {Mann}, A.~W., {et~al.} 2020, \aj, 160, 33,
  \dodoi{10.3847/1538-3881/ab94b7}

\bibitem[{{Shoda} {et~al.}(2024){Shoda}, {Namekata}, \& {Takasao}}]{shoda24-1}
{Shoda}, M., {Namekata}, K., \& {Takasao}, S. 2024, \aap, 691, A152,
  \dodoi{10.1051/0004-6361/202450129}

\bibitem[{{Thao} {et~al.}(2024){Thao}, {Mann}, {Feinstein}, {Gao}, {Thorngren},
  {Rotman}, {Welbanks}, {Brown}, {Duvvuri}, {France}, {Longo}, {Sandoval},
  {Schneider}, {Wilson}, {Youngblood}, {Vanderburg}, {Barber}, {Wood},
  {Batalha}, {Kraus}, {Murray}, {Newton}, {Rizzuto}, {Tofflemire}, {Tsai},
  {Bean}, {Berta-Thompson}, {Evans-Soma}, {Froning}, {Kempton}, {Miguel}, \&
  {Pineda}}]{thao24-1}
{Thao}, P.~C., {Mann}, A.~W., {Feinstein}, A.~D., {et~al.} 2024, \aj, 168, 297,
  \dodoi{10.3847/1538-3881/ad81d7}

\bibitem[{{Tian} {et~al.}(2008){Tian}, {Kasting}, {Liu}, \& {Roble}}]{tian08-1}
{Tian}, F., {Kasting}, J.~F., {Liu}, H.-L., \& {Roble}, R.~G. 2008, Journal of
  Geophysical Research (Planets), 113, E05008, \dodoi{10.1029/2007JE002946}

\bibitem[{{Tu} {et~al.}(2015){Tu}, {Johnstone}, {G{\"u}del}, \&
  {Lammer}}]{tu15-1}
{Tu}, L., {Johnstone}, C.~P., {G{\"u}del}, M., \& {Lammer}, H. 2015, \aap, 577,
  L3, \dodoi{10.1051/0004-6361/201526146}

\bibitem[{Virtanen {et~al.}(2020)Virtanen, Gommers, Oliphant, Haberland, Reddy,
  Cournapeau, Burovski, Peterson, Weckesser, Bright, {van der Walt}, Brett,
  Wilson, Millman, Mayorov, Nelson, Jones, Kern, Larson, Carey, Polat, Feng,
  Moore, {VanderPlas}, Laxalde, Perktold, Cimrman, Henriksen, Quintero, Harris,
  Archibald, Ribeiro, Pedregosa, {van Mulbregt}, \& {SciPy 1.0
  Contributors}}]{virtanenetal20-1}
Virtanen, P., Gommers, R., Oliphant, T.~E., {et~al.} 2020, Nature Methods, 17,
  261, \dodoi{10.1038/s41592-019-0686-2}

\bibitem[{{Wilson} {et~al.}(2022){Wilson}, {Youngblood}, {Toloza}, {Drake},
  {France}, {Froning}, {G{\"a}nsicke}, {Redfield}, \& {Wood}}]{wilson22-1}
{Wilson}, D.~J., {Youngblood}, A., {Toloza}, O., {et~al.} 2022, \apj, 936, 189,
  \dodoi{10.3847/1538-4357/ac87a8}

\bibitem[{{Wood} {et~al.}(2005){Wood}, {Redfield}, {Linsky}, {M{\"u}ller}, \&
  {Zank}}]{woodetal05-1}
{Wood}, B.~E., {Redfield}, S., {Linsky}, J.~L., {M{\"u}ller}, H.-R., \& {Zank},
  G.~P. 2005, \apjs, 159, 118, \dodoi{10.1086/430523}

\bibitem[{{Woodgate} {et~al.}(1998){Woodgate}, {Kimble}, {Bowers}, {Kraemer},
  {Kaiser}, {Danks}, {Grady}, {Loiacono}, {Brumfield}, {Feinberg}, {Gull},
  {Heap}, {Maran}, {Lindler}, {Hood}, {Meyer}, {Vanhouten}, {Argabright},
  {Franka}, {Bybee}, {Dorn}, {Bottema}, {Woodruff}, {Michika}, {Sullivan},
  {Hetlinger}, {Ludtke}, {Stocker}, {Delamere}, {Rose}, {Becker}, {Garner},
  {Timothy}, {Blouke}, {Joseph}, {Hartig}, {Green}, {Jenkins}, {Linsky},
  {Hutchings}, {Moos}, {Boggess}, {Roesler}, \& {Weistrop}}]{woodgateetal98-1}
{Woodgate}, B.~E., {Kimble}, R.~A., {Bowers}, C.~W., {et~al.} 1998, \pasp, 110,
  1183, \dodoi{10.1086/316243}

\bibitem[{{Woods} {et~al.}(2009){Woods}, {Chamberlin}, {Harder}, {Hock},
  {Snow}, {Eparvier}, {Fontenla}, {McClintock}, \& {Richard}}]{woods09-1}
{Woods}, T.~N., {Chamberlin}, P.~C., {Harder}, J.~W., {et~al.} 2009, \grl, 36,
  L01101, \dodoi{10.1029/2008GL036373}

\bibitem[{{Wright} {et~al.}(2011){Wright}, {Drake}, {Mamajek}, \&
  {Henry}}]{wright11-1}
{Wright}, N.~J., {Drake}, J.~J., {Mamajek}, E.~E., \& {Henry}, G.~W. 2011,
  \apj, 743, 48, \dodoi{10.1088/0004-637X/743/1/48}

\bibitem[{{Youngblood} {et~al.}(2022){Youngblood}, {Pineda}, {Ayres}, {France},
  {Linsky}, {Wood}, {Redfield}, \& {Schlieder}}]{youngbloodetal22-1}
{Youngblood}, A., {Pineda}, J.~S., {Ayres}, T., {et~al.} 2022, arXiv e-prints,
  arXiv:2201.01315.
\newblock \doarXiv{2201.01315}

\bibitem[{{Youngblood} {et~al.}(2021){Youngblood}, {Pineda}, \&
  {France}}]{youngbloodetal21-1}
{Youngblood}, A., {Pineda}, J.~S., \& {France}, K. 2021, \apj, 911, 112,
  \dodoi{10.3847/1538-4357/abe8d8}

\bibitem[{{Zhu} \& {Preibisch}(2025)}]{zhu25-1}
{Zhu}, E., \& {Preibisch}, T. 2025, \aap, 694, A93,
  \dodoi{10.1051/0004-6361/202452057}

\end{thebibliography}

\end{document}